\documentclass[10pt,aps,twocolumn,superscriptaddress,prb,longbibliography]{revtex4-2}

\usepackage[utf8]{inputenc}
\usepackage[T1]{fontenc}
\usepackage{dcolumn}
\usepackage{bm}
\usepackage{amsmath, amssymb}
\usepackage{natbib}
\usepackage{graphicx} 
\usepackage{tabularx}
\usepackage{multirow}
\usepackage{rotating} 
\usepackage{makecell}
\usepackage{parskip}
\usepackage{textgreek}
\usepackage[normalem]{ulem}
\usepackage{hyperref}
\hypersetup{colorlinks=true, linkcolor=blue, citecolor=blue, urlcolor=blue}
\usepackage{float}
\usepackage[english]{babel}
\usepackage{tabularx}
\usepackage{booktabs}
\usepackage{lipsum}
\usepackage{graphicx}
\usepackage{siunitx}
\usepackage{bbold}

\newcommand{\cm}{~cm$^{-1}$~}

\usepackage{lineno}

\begin{document}

\title{Coupling of magnetic and lattice collective excitations in the 2D van der Waals antiferromagnet FePS$_3$}

\author{Kartik Panda}
\affiliation{Department of Physics, Ariel University, Ariel 40700, Israel}

\author{Itzik Kapon}
\affiliation{Department of Quantum Matter Physics, University of Geneva, CH-1211 Geneva 4, Switzerland}

\author{Dumitru Dumcenco}
\affiliation{Department of Quantum Matter Physics, University of Geneva, CH-1211 Geneva 4, Switzerland}

\author{Dirk van der Marel}
\affiliation{Department of Quantum Matter Physics, University of Geneva, CH-1211 Geneva 4, Switzerland}

\author{Alexey Kuzmenko}
\affiliation{Department of Quantum Matter Physics, University of Geneva, CH-1211 Geneva 4, Switzerland}

\author{Nimrod Bachar}
\email{nimib@ariel.ac.il}
\affiliation{Department of Physics, Ariel University, Ariel 40700, Israel}

%\date{\today}

\begin{abstract}
We combine polarized infrared magneto-transmission and Faraday angle rotation measurements to map the collective excitations of the van der Waals antiferromagnet FePS$_3$. Below the N\'{e}el temperature ($T_\mathrm{N} \approx 118~\mathrm{K}$), the phonon spectrum becomes strongly anisotropic, reflecting the underlying zigzag antiferromagnetic order. In contrast, a prominent excitation at $122~\mathrm{cm}^{-1}$ ($15$~meV) is polarization-independent, hardens on cooling, and splits linearly with magnetic field, identifying its magnetic origin. From absolute transmission and Faraday rotation, we reconstruct the circular optical conductivities and reveal a pronounced dichroism of the field-split excitations. The upper branch near $129~\mathrm{cm}^{-1}$ exhibits a reduced dichroic response, consistent with hybridization with a nearby infrared phonon. Several phonon modes exhibit sizable Faraday rotation, providing evidence for spin-phonon coupling and demonstrating that lattice vibrations acquire magnetic-field-dependent optical activity. In addition, additional excitations appear in the infrared spectra and a broad mid-infrared feature near $900~\mathrm{cm}^{-1}$ emerges only below $T_\mathrm{N}$, consistent with a modified lattice response in the magnetic state. These results highlight the anisotropic nature of spin--phonon coupling in FePS$_3$ and establish polarization-resolved magneto-optical spectroscopy as a powerful probe of coupled spin and lattice dynamics in two-dimensional antiferromagnets.
\end{abstract}

%\pacs{nn.nn.xx}

%\keywords{}

\maketitle

\section{\label{sec:intro}Introduction}

In recent years, atomically thin crystals, from graphene to transition-metal dichalcogenides, have transformed condensed-matter research and inspired a wide range of device concepts~\cite{Geim2009,Chhowalla2013}. Beyond their celebrated electronic properties, an expanding family of two-dimensional (2D) compounds displays intrinsic magnetism, unlocking fresh opportunities for spintronic and information-processing technologies~\cite{Ahn2020,Nmec2018}. Consequently, major efforts have focused on reliable exfoliation, scalable synthesis, and device integration of 2D magnets. Despite these advances, fundamental questions remain about how magnetic order couples to lattice dynamics in 2D antiferromagnets, and whether unique optical signatures of spin–lattice interactions can be harnessed to probe or control magnetic phases.

Two-dimensional magnets are also an unparalleled platform for testing the fundamental physics of spin models. They permit direct access to Ising, XY, Kitaev, and Heisenberg-type physics within a single atomic layer~\cite{Gibertini2019}. This enables direct observation and control of magnetic fluctuations, and allows one to probe how external stimuli such as temperature, magnetic field, strain, and pressure modify the ground state~\cite{Burch2018,Huang2017,Gong2017}. Robust (anti)ferromagnetic order in the two-dimensional limit can emerge in the presence of magnetic anisotropy, interlayer coupling, or engineered perturbations, providing a highly tunable platform for exploring correlated spin phenomena.

Magnetic van der Waals materials, including metal phosphorus trichalcogenides (MPX$_{3}$) [M=Fe, Co, Ni, Mn; X=S, Se], represent a broad class of compounds with diverse spin dimensionality models, exhibiting both ferromagnetic and antiferromagnetic states. For example, FePS$_3$ ($T_\mathrm{N} \approx 120$ K), MnPS$_{3}$ ($T_\mathrm{N} \approx 78$ K), and NiPS$_{3}$ ($T_\mathrm{N} \approx 155$ K), all with monoclinic crystal structures (space group C2/m), display distinct magnetic behaviors~\cite{Brec1986,Joy1992a}. FePS$_{3}$ adheres to an Ising model with a zigzag antiferromagnetic configuration, featuring in-plane ferromagnetic chains that are antiparallel along one of the in-plane crystal axes, while NiPS$_{3}$ and MnPS$_{3}$ are best described by an XY and Heisenberg model, respectively~\cite{Brec1986,Joy1992a}. MnPS$_3$ shows a N\'{e}el-type antiferromagnetic order, where neighboring atoms have antiparallel spins, in contrast to the zigzag antiferromagnetic structure in FePS$_3$. Notably, FePS$_3$ maintains its magnetic order at the monolayer limit~\cite{Lee2016,Wang2016} and undergoes an insulator-to-metal transition under quasi-hydrostatic pressure~\cite{Mallick2024}. Inelastic neutron scattering measurements of FePS$_{3}$ have identified two excitations with finite gaps of 15 meV and 40 meV at $q = 0$~\cite{Lanon2016}. Moreover, the comparable magnitudes of spin-orbit coupling and crystal field energies in FePS$_3$ favor the presence of low-energy magnetic excitations~\cite{Yuan2024}. Therefore, FePS$_3$ provides a model system for studying the interplay between anisotropic magnetism and lattice dynamics in van der Waals antiferromagnets. Yet, a comprehensive understanding of how these low-energy magnetic excitations interact with phonons, what their circular-conductivity spectral weight is, and how they affect crystal symmetries in FePS$_3$ remains lacking.

The magnetic properties of these vdW materials are strongly coupled to the charge and lattice degrees of freedom and can be studied effectively using far-infrared spectroscopy. The advantage of optical spectroscopy is that it is also susceptible to the monolayer limit. Therefore, the magnetic state can be easily studied and detected using optical probes, such as Kerr/Faraday angle rotation, compared to other magnetometry techniques~\cite{Mak2019,Shree2020}. The additional advantage of infrared spectroscopy is direct access to the spectral weight of the different collective modes in the magnetic phase and, as a result, to their coupling and their relation to the underlying physics of the magnetic state.     

In this work, we investigate the low-energy collective excitations of FePS$_3$ using polarization-resolved infrared magneto-transmission and Faraday rotation measurements. Our goal is to elucidate the interplay between lattice vibrations and magnetic excitations in the antiferromagnetic state, and to determine how the underlying zigzag spin structure influences the optical response. By combining polarized spectroscopy with quantitative magneto-optical analysis, we access the circular optical conductivity and identify signatures of anisotropic spin--phonon coupling. In particular, we provide a polarization-resolved characterization of the phonon spectrum, and report Faraday rotation measurements of the magnetic excitations at 122~\cm\ and 320~\cm, enabling a quantitative determination of their circular optical response. This approach provides new insight into the coupling between spin and lattice degrees of freedom in low-dimensional antiferromagnets.

\section{Methods\label{sec:methods}}

\subsection{Sample preparation}\label{sec:sample}

FePS$_3$ crystals were grown by the Chemical Vapor Transport (CVT) method, like other magnetic van der Waals materials in the MPX3 group~\cite{DUMCENCO2020125799}. A stoichiometric amount in the ratio 1:1:3 of elements Fe (99.9\%, Chempure), P (99.99\%, Alfa Aesar), and S (99.999\%, Fluka), corresponding to approximately 0.5 gram total mass, was inserted in a quartz tube inside a glove box, together with approximately 2 mg/cm\(^{3}\) of iodine as transport agent. The tube was subsequently evacuated down to ~10\(^{-4}\) mbar, with intermediate Ar flushing, and sealed to a length of 16 cm. The sealed tube was inserted in a horizontal furnace, in a controlled temperature gradient with 680~$^{\circ}$C and 630~$^{\circ}$C for the hot and cold ends, respectively. At the end of 20 days, the furnace was switched off, and the crystals were extracted from the tube.

\subsection{Infrared spectroscopy}\label{sec:optics}

To measure the magneto-optical coefficients, we have used a Bruker Vertex 70v Fourier Transform Infra-Red (FTIR) spectrometer coupled to a Cryogenic magneto-optical cryostat, allowing transmission measurements as a function of temperature, magnetic field, and incoming/outgoing polarization. We have used Hg arc lamp and Globar (SiC) sources for the far-infrared (FIR) and mid-infrared (MIR) frequency ranges, respectively. Residual magnetic fields affect the arc lamp, limiting reliable measurements below 40\cm. Therefore, we primarily present results obtained with the Globar source. To detect the transmitted signal intensity in our measurements, we have used an IR Labs liquid-helium-cooled (and pumped) Si bolometer with cutoff filter for the range of 10 to 80 \cm, an IR Labs liquid-helium-cooled Si bolometer with a cutoff filter for the range of 40 to 700 \cm, and an IR Labs liquid-helium-cooled Si bolometer with three different cutoff filters covering the range of 500 to 4000 \cm. The magnetic field of the cryostat can be tuned from -7~T to +7~T and was aligned parallel (perpendicular) to the light propagation vector (sample surface normal) in the Faraday configuration. The temperature range of the cryostat sample holder is from 4~K to about 300~K, while the sample is mounted with silver paste to the sample holder and situated in the vacuum chamber as the superconducting magnet during the measurement. The cryostat is positioned on an external linear translation stage, allowing the measurement of a sample in transmission (reflection) configuration and a bare hole (mirror) reference at any desired temperature, magnetic field, and incoming/outgoing polarization. 

Three types of experiments were done in this work: (i) Polarization-dependent transmission measurements using gold or KRS-5 (thallium bromoiodide) polarizers for the FIR or MIR frequency ranges, respectively, along the incoming and outgoing light beam paths. Our preliminary attempts to measure and fit non-polarized transmission data and, as a result, to extract the optical conductivity have failed, indicating the importance of the polarization dependence of the transmission spectrum and the anisotropy of the monoclinic crystal of FePS$_3$ in the magnetic state. Therefore, we have used an incoming polarizer for the transmission measurements. As a first stage, we measured the FIR and MIR spectra as a function of polarizer angle deep in the magnetic state, at 5~K, and at zero magnetic field (See App.~\ref{sec:appTr}). We have identified two distinct transmission spectra separated by a $90^{\circ}$ polarizer angle difference, to which we refer here as "Pol 1" and "Pol 2". We emphasize that these two principal polarizations were not associated with the $a-b$ plane crystal axis using the X-ray diffraction apparatus and are only determined by the optical transmission spectra. Following this criterion, we have used these two principal polarizations for all the temperature and magnetic field-dependent transmission measurements without an outgoing analyzer. (ii) Faraday rotation angle measurements with a rotating polarizer at the incoming optical path and a fixed analyzer at the outgoing optical beam. The fast protocol was used to investigate the temperature and magnetic field evolution of the Faraday angle rotation in FePS$_3$. Since FePS$_3$ has a monoclinic crystal structure and shows an anisotropic optical response in the antiferromagnetic state, the typical fast protocol procedure~\cite{Levallois2015}, taking into account only four measurements, i.e. combination of external magnetic field $\pm B$ and $A = P = \pm 45^{\circ}$, where $P$ and $A$ are the incoming polarizer and outgoing analyzer angles, respectively, was subject to possible birefringence artifacts. Therefore, we have conducted (iii) full protocol Faraday rotation angle measurements for each of the two principal polarizations used in this work. In the full protocol, the analyzer angle is fixed to one of the principal angles while measurement of the transmission intensity is being taken at intervals of $15^{\circ}$ for a full $360^{\circ}$ polarizer rotation. We have conducted the full protocol at a temperature of 5~K, i.e., deep in the antiferromagnetic state, and at fields of $B=\pm 7~T$.  

\subsection{Raman microscopy}\label{sec:Raman}

Several samples of FePS$_3$ were measured in Raman spectroscopy using a LabRAM HR Evolution Horiba Confocal Raman Microscope system combined with a CryoVac liquid helium flow cryostat. A 532-nm unpolarized laser was used as the excitation source, focused on the sample (inside the cryostat) through a 63× Olympus objective, and the laser power was set to 0.4 mW throughout the experiment. The scattered light from the sample was collected by the same objective, passed through the analyzer, and sent to a Czerny-Turner spectrometer equipped with either a 600- or 1800-grooves-per-mm grating, and was detected by a liquid nitrogen-cooled CCD array. The spectra were obtained as a function of temperature from 300~K down to 10~K at 10~K intervals, in addition to a measurement at 5~K, which is the base temperature of the Cryostat in our system. The 1800 grooves per mm grating was used to measure the Raman shift spectrum in the range of about $\pm$600\cm of Raman shift (i.e. Stokes and anti-Stokes transitions) around the elastic line and to resolve the frequency resolution of the different excitations, in particular the modes around 80 to 130\cm appearing below the phase transition of FePS$_3$. The 600 grooves per mm grating was used to measure the large Raman shift range of up to approximately 4000\cm in order to resolve the excitations seen in our data and in previous reports on FePS$_3$ around 1000\cm (e.g, Ref.~\citep{Jouanne1989}).      

\subsection{Phonon calculation}\label{sec:DFT}

In order to obtain the phonon frequencies of FePS$_3$, we have performed first-principles Density-functional theory (DFT) calculations with the Quantum ESPRESSO PWscf (Plane-Wave Self-Consistent Field) package v6.6~\cite{Giannozzi2009,Giannozzi2017,QEwebsite}. Exchange–correlation effects were treated within the PBE (Perdew-Burke-Ernzerhof) functional and the generalized-gradient approximation (GGA)~\cite{Perdew1996}. Scalar-relativistic pseudopotentials (PPs) consistent with PBE were used: PAW (Projector-Augmented Wave) for Fe and US (Ultrasoft) for P and S~\cite{Prandini2018}. The lattice parameters were adopted from the parameters of FePS$_3$ in the Materials Project database~\cite{Jain2013,MatProjFPS}. Wavefunctions and charge density were expanded with kinetic-energy cutoffs of 90 Ry and 1080 Ry, respectively. Metallic occupations were treated using Marzari–Vanderbilt cold smearing with a width of 0.0147 Ry ($\approx$ 0.20 eV)~\cite{Marzari1999}. Brillouin-zone sampling employed a Monkhorst–Pack $6\times4\times3$ mesh~\cite{Monkhorst1976}. Symmetry was enabled, and four symmetry operations (including inversion, two with fractional translations) were detected. Electronic self-consistency used plain mixing with a mixing parameter $\beta=0.4$ and a target energy change $ \mathrm{conv_{thr}}=8\times10^{-9}$ Ry; in this calculation, convergence was reached in 20 SCF iterations.

\begin{figure}
    \centering
    \includegraphics[width=0.9\linewidth]{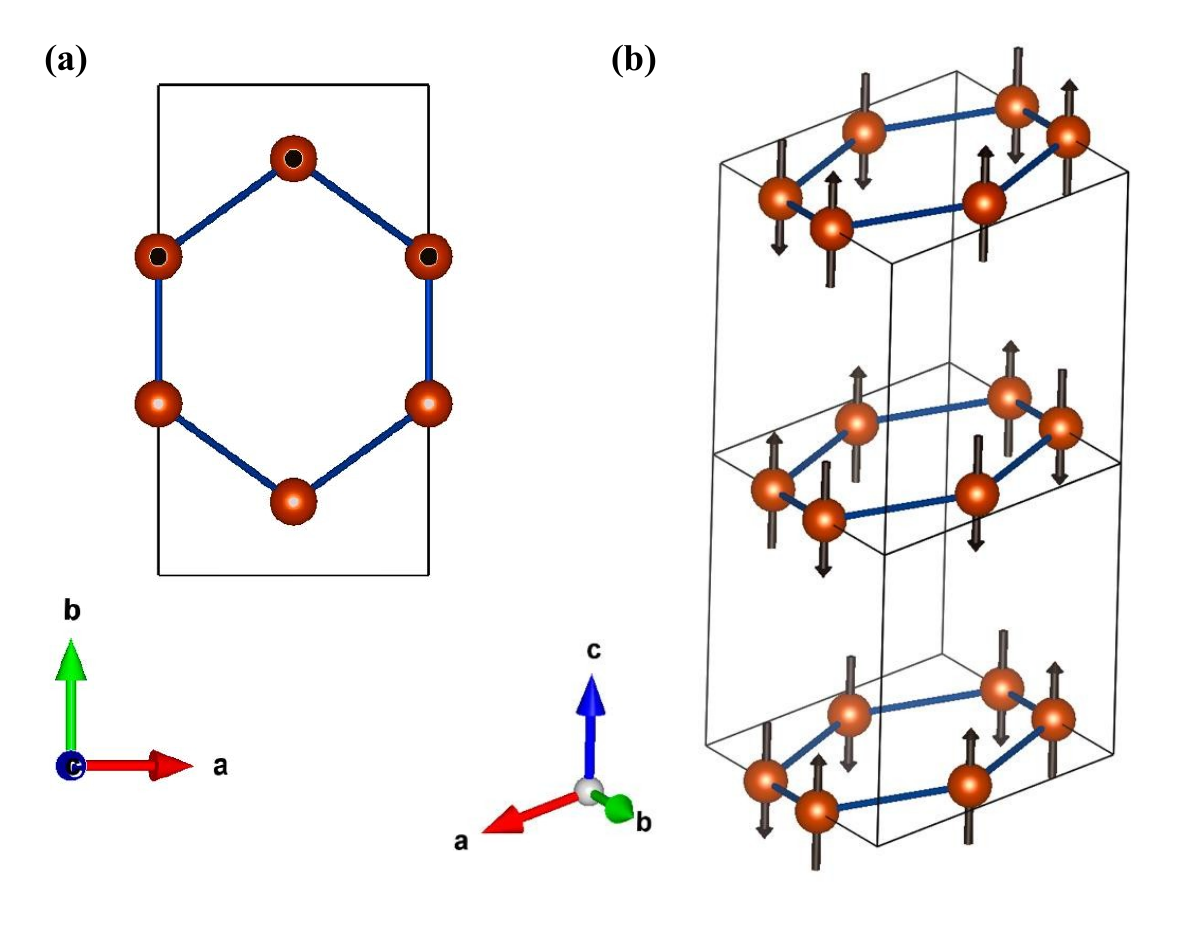}
       \caption{Illustration of the crystal structure of FePS$_3$. (a) Top view of the Fe honeycomb lattice in the $ab$ plane. (b) Three-dimensional view of the crystal structure, showing the zigzag antiferromagnetic order with ferromagnetic spin chains along the $a$-axis, coupled antiferromagnetically along the $b$-axis and between adjacent layers along the $c$-axis. The spheres with black (white) dots represent magnetic moments oriented parallel (antiparallel) to the $c$-axis.}
    \label{fig:FPSCrystal}
\end{figure}

FePS$_3$ crystallizes in the monoclinic space group $C2/m$ (point group $C_{2h}$) of the paramagnetic state like other members of the magnetic van der Waals MPX$_3$ family. Therefore the the optical phonons at the $\Gamma$ point decompose as $\Gamma = 8A_g + 7B_g + 6A_u + 9B_u$. The $A_g$ and $B_g$ modes are Raman-active, while $A_u$ and $B_u$ modes are infrared-active. Since FePS$_3$ magnetic state is composed of zigzag chains along the $a$-axis, which are coupled antiferromagnetically to their neighbors along the $b$-axis and between the crystal planes along the $c$-axis (See Figure~\ref{fig:FPSCrystal}), we have tried to capture this structure with a collinear spin polarization ($n_{\mathrm{spin}}=2$) of two Fe sublattices in the $a-b$ plane initialized antiparallel (Fe1/Fe2 starting magnetization $\pm0.3125$), and small seed magnetizations on P and S (0.1). As a result, calculations were performed for 20 atoms per cell across four atomic species. The converged self-consistent solution has essentially zero net moment with total magnetization $\approx 0~\mu_{\mathrm{B}}$ per cell (absolute magnetization $\sim15~\mu_{\mathrm{B}}$). Phonons were computed within density-functional theory (DFT) as implemented in QE’s ph.x~\cite{Baroni2001} and resulted in zone-center ($\Gamma$-point) phonon frequencies with twice the number of modes compared to the conventional paramagnetic cell calculation. The symmetry assignments used in Table~\ref{tab:IRphononlist} are based on polarization-resolved infrared measurements and compared with DFT calculations. Experimental values for our Raman measurements, along with their assignment to our DFT calculation, are given in Table~\ref{tab:Rphononlist} of the appendix. 

% Pol1 = MIR Pol1=200 and FIR Pol2=150
% Pol2 = MIR Pol2=290 and FIR Pol1=60

\section{Results\label{sec:results}}

\subsection{Optical Conductivity}

\begin{figure}
    \centering
    \includegraphics[width=\linewidth]{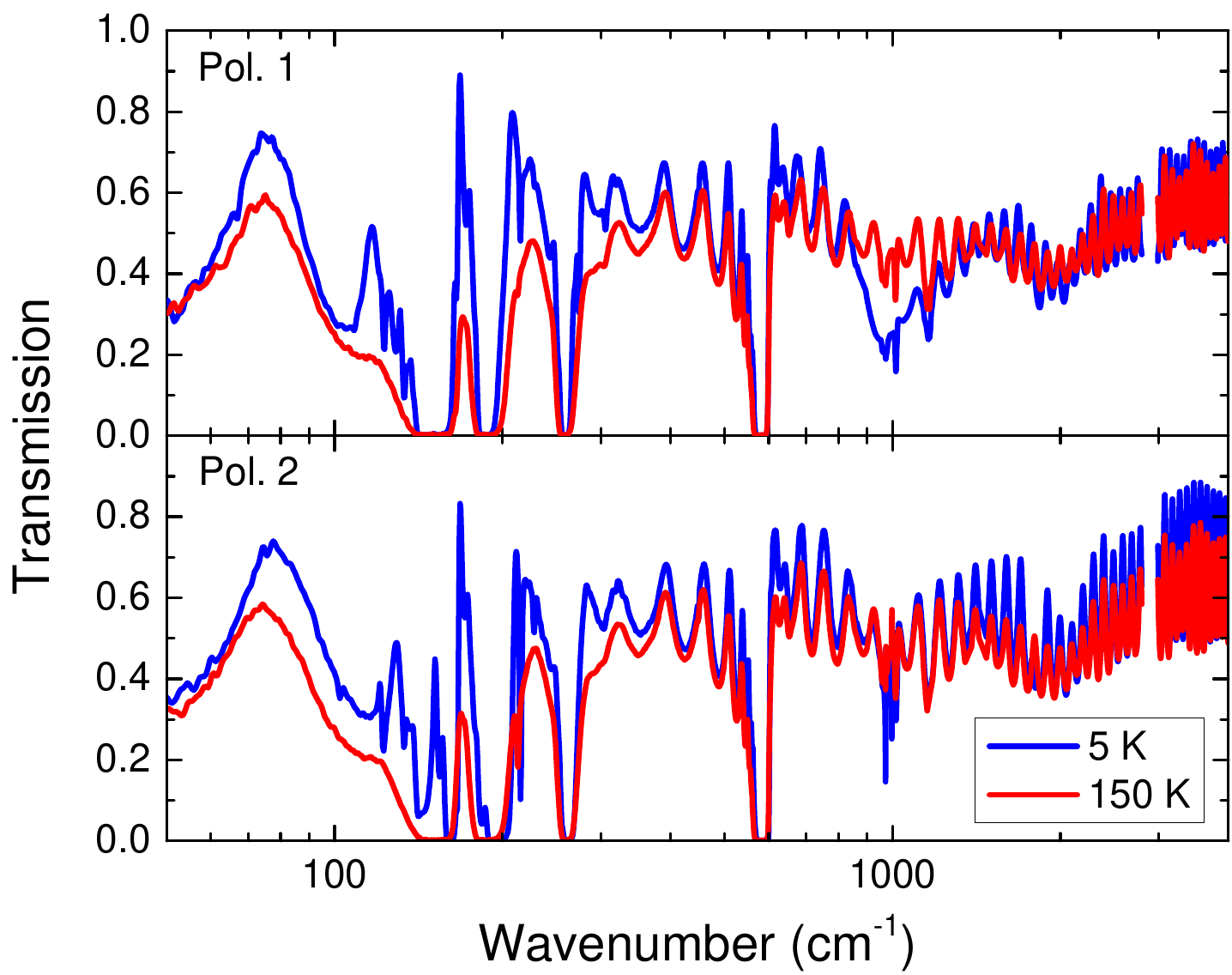}
       \caption{Polarized transmission through a 26~$\mu$m FePS$_3$ flake as a function of frequency (in log scale) at selected temperatures (Blue line: $5~K \ll T_N$, Red line: $150~K > T_N$) for two principal polarizations (upper panel: Polarization 1, and lower panel: Polarization 2. See Sec .~\ref{sec:optics} for definition). The logarithmic scale was chosen to present the entire dataset from the FIR to the MIR range. A pronounced spectral change occurs below $T_N = 118$~K, with the emergence of new absorption lines in the range of 50 to 250\cm in both polarizations and an absorption feature around 900\cm in only one of the polarizations. The full temperature-dependent dataset of the transmission spectra is available via an open-access data repository~\cite{OpenData2026}.}
    \label{fig:ZF_T_Compare}
\end{figure}

The polarized transmission and the polarized optical conductivity results in this work are reported here for the first time for FePS$_3$ and will allow us to disentangle and tentatively associate the collective excitation spectra with the phonons and crystal symmetries in the magnetic state. We start by noting that the transmission spectra at temperatures substantially higher than the N\'{e}el antiferromagnetic transition temperature, $T_N=118~K$, of FePS$_3$ show no major difference between the two principal polarization measurement axes (See in App.~\ref{sec:appTr} for the comparison) as can be seen in Figure~\ref{fig:ZF_T_Compare}. Below $T_N$, the transmission spectra exhibit several new absorption lines in the low frequency range ($<200~cm^{-1}$), among them is a mode at 122\cm (15~meV), in both polarizations, and a wide absorption feature at around 900\cm appearing only in the 1st polarization, as shown in Fig.~\ref{fig:ZF_T_Compare}. 

\begin{figure*}
    \centering
    \includegraphics[width=1.0\linewidth]{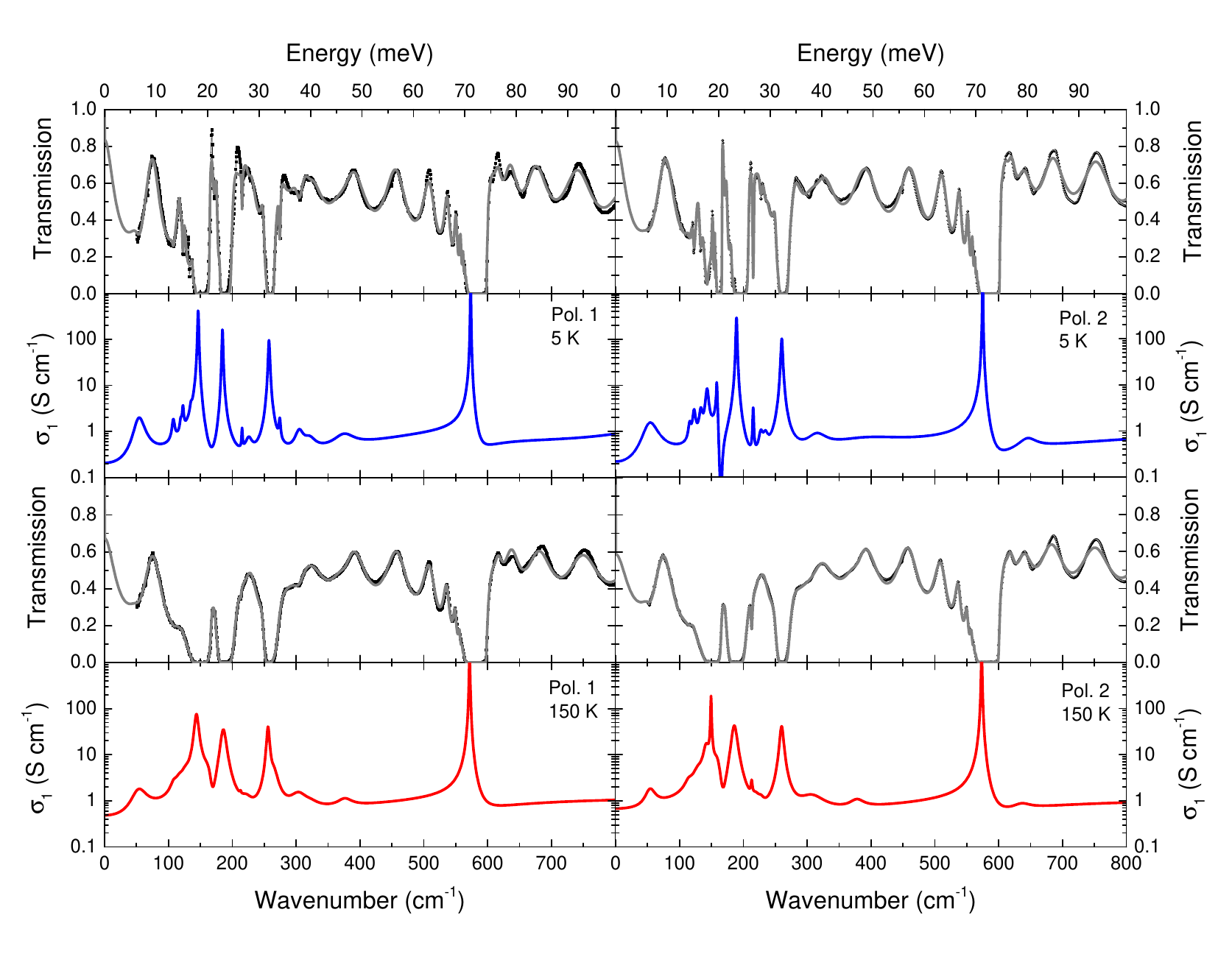}
      \caption{Transmission spectra in the FIR range (black symbols) along with the Drude-Lorentz fit (gray line) from which the real part of the optical conductivity, $\sigma_1$, was extracted at temperatures of $5~K \ll T_N$ (Top panels) and $150~K > T_N$ (lower panels) and zero magnetic field for the two principal polarizations.}
    \label{fig:Tr_Sigma_Fit}
\end{figure*}

The transmission spectra are fitted in the range of 40 to 4000\cm with a plane-parallel slab transmission model in which the slab properties are defined by the Drude-Lorentz model:
\begin{equation}
    \varepsilon(\omega)=\varepsilon_\infty+\sum_j\frac{\omega^2_{p_j}}{\omega^2_{0_j}-\omega^2-i\gamma_j\omega}
\end{equation}
from which we obtain the central frequency $\omega_0$, plasma frequency $\omega_p$, and width $\gamma$ of each mode $j$. Here $\varepsilon_\infty$ is the high-frequency dielectric constant. We note that this analysis assumes $\mu \approx 1$. This is a standard approximation in infrared studies of magnetic semiconductors, where magnetic permeability contributions are typically small compared to the dielectric response in the investigated frequency range. A more complete treatment, including magnetic permeability effects, is left for future work. The Drude-Lorentz fit is then applied to all transmission spectra measured in this work from 150~K down to 5~K. Figure~\ref{fig:Tr_Sigma_Fit} shows a typical example of the fit (gray line) at selected temperatures for each polarization. The real part of the optical conductivity, $\sigma_1$, of the slab resulting from the Drude-Lorentz fit is shown in Figure~\ref{fig:Tr_Sigma_Fit}, from which we can already note some of the polarization-dependent modes of FePS$_3$. 
At 150~K we detect four strong modes for each polarization situated at 144, 186.1, 256.1, and 571.8\cm for the 1st polarization and 149.5, 185, 260.1, and 573.4\cm for the 2nd polarization, where the values are taken from the Drude-Lorentz fit to the transmission data. In addition to the main modes, there are sidebands around 160\cm and below 140\cm in both polarizations, which are in the frequency range where new modes are detected in the low-temperature spectra below T$_N$. 

The change in the transmission spectra across the N\'{e}el transition is attributed to several new modes which mainly affect the low frequency optical conductivity. Figure~\ref{fig:Tr_Sigma_Fit} shows some of these modes, which can be captured directly by the Drude-Lorentz fit, for example, the mode at 122\cm, seen in both polarizations. The strong modes at 184.7, 257.6, and 573.4\cm for the 1st polarization and 189.4, 260.4, and 574.9\cm for the 2nd polarization are associated with the modes above T$_N$, mentioned before, due to their close frequencies. An important anisotropy feature in the antiferromagnetic state shows up when the 147.3\cm mode is observed as a strong IR peak in the 1st polarization, while the 149.5\cm mode in the 2nd polarization shows a substantial decrease in the optical conductivity by about one order of magnitude compared to its oscillator strength above T$_N$. Moreover, the mode at 215\cm, observed above the phase transition, is mainly active in the 2nd polarization and is further enhanced in the antiferromagnetic state.

The Drude-Lorenz fit is, however, not capturing all the fine details of the transmission spectra, in particular in the low frequency range of our measurements, and close to most of the new spectral features appearing below T$_N$. This is particularly evident in the absorption lines at 128 and 133\cm in the 1st polarization, and at about 158\cm in the 2nd polarization, where our attempts to capture the frequency dispersion details with the Drude-Lorentz model were unsuccessful. Therefore, further to the fit process, we have employed the Variable Dielectric Function (VDF) method~\cite{Kuzmenko2005} to the raw transmission data for both principal polarizations and at all measured temperatures from 150~K down to 5~K in order to fully capture and investigate the temperature and polarization dependent optical conductivity of our FePS$_3$ sample.

\begin{figure}
    \centering
    \includegraphics[width=1.0\linewidth]{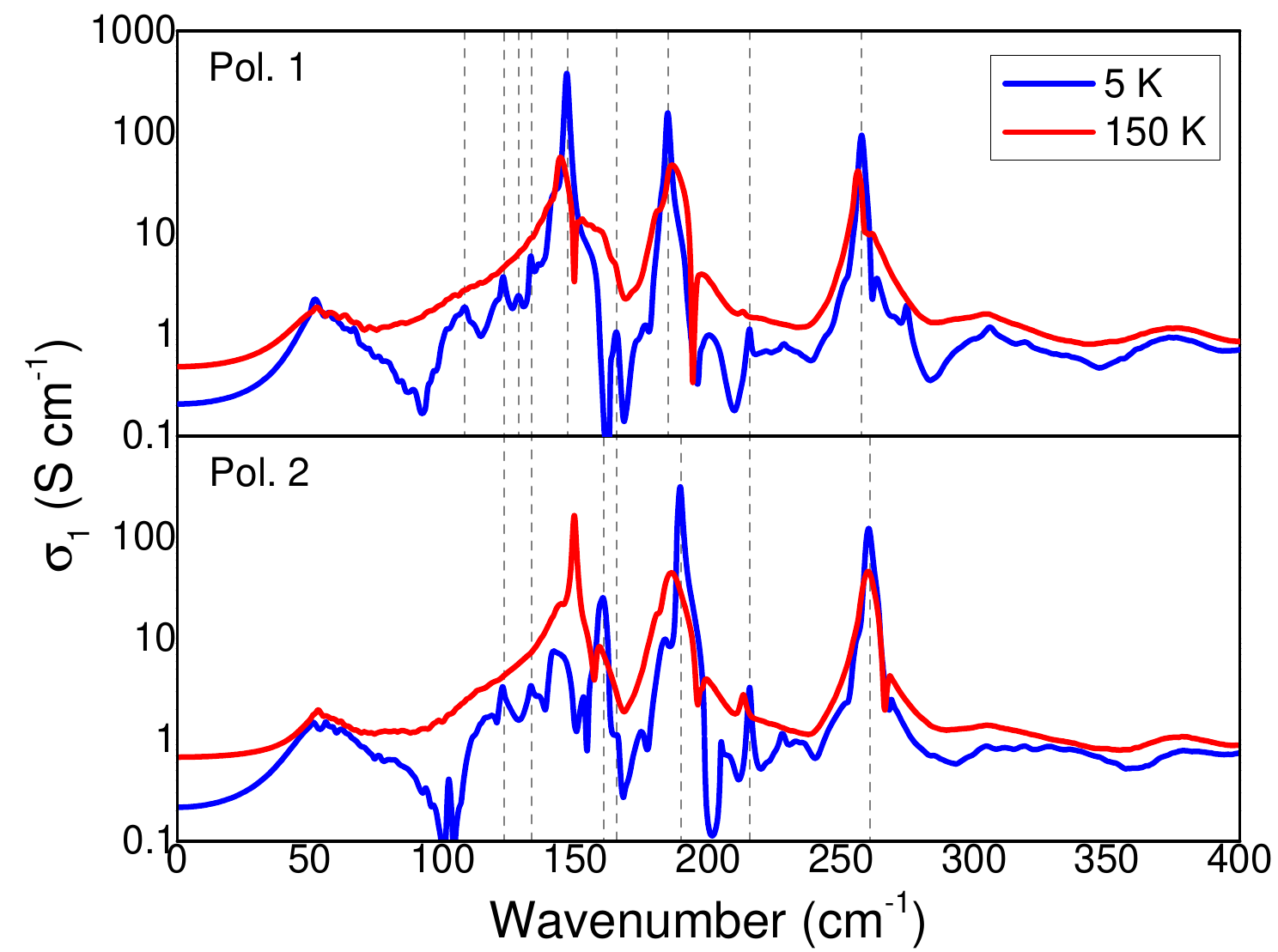}
      \caption{Real part of the optical conductivity $\sigma_1$ at 150~K and 5~K in zero magnetic field, extracted via the Variable Dielectric Function of the Drude–Lorentz modeling of transmission data for the two principal polarizations.}
    \label{fig:Sigma1}
\end{figure}

Figure~\ref{fig:Sigma1} compares the optical conductivity after employing the VDF method for each polarization at the selected temperatures of 150~K ($>T_N$) and 5~K ($\ll T_N$), focusing mainly on the low-frequency range where new optical modes appear in the data of the antiferromagnetic state. Additional comparison across the entire frequency range of our work is provided in the appendix, along with a full repository of the optical conductivity dataset for all temperatures~\cite{OpenData2026}.  
The transmission data and the real part of the optical conductivity reveal anisotropy in the phonon spectra, emerging below the antiferromagnetic phase transition, particularly in the low-energy range for the two principal polarizations. As noted above, the strong phonon mode at 146.7~\cm\ in Pol.~1 exhibits a significantly reduced spectral weight in Pol.~2. Similarly, the IR mode at 128~\cm\ is prominent in Pol.~1, while a distinct mode at 160~\cm\ appears only in Pol.~2. We present in Table~\ref{tab:IRphononlist} the full list of modes obtained at 5~K. The modes discussed in the text are indicated by dashed vertical lines in Fig.~\ref{fig:Sigma1} and highlighted in bold in Table~\ref{tab:IRphononlist}. We will elaborate on their assignments in the following section; however, we emphasize that the symmetry assignments are indicative, based on the closest correspondence between experimental frequencies and DFT calculations, and should not be interpreted as a rigorous symmetry classification.

\begin{table*}
\centering
\caption{Experimentally observed infrared excitations in FePS$_3$ compared with DFT phonon calculations performed in this work. Modes are reported and assigned by their polarizations (Pol.~1 and Pol.~2). Strong modes are considered as modes with optical conductivity values of the order of 100~$\Omega^{-1} cm^{-1}$ and above. DFT phonon symmetries and activities are taken from the calculated phonon spectrum and assigned based on the discussion in the main text and their matching frequency in the experimental data. All values are in cm$^{-1}$. Modes in bold text are those marked in dashed lines in Figure~\ref{fig:Sigma1}. In the comment, we also cite modes previously reported in non-polarized studies.}
\begin{tabular}{|c|c|c|c|}
\hline
\textbf{Exp. freq.} & \textbf{IR Pol.}  & \textbf{DFT freq.} & \textbf{Comment} \\
\hline

52      &  1,2   & --               & Wide IR feature; Ref.~\citep{Bernasconi1988} \\
\hline

102.3   &  2     & 101.9 ($B_u$)    &  \\
\hline

\textbf{108.1}   &  1     & --               & Wide IR feature; \\
        &        &                  & Shows Faraday angle \\
\hline

119.5   &  1     & --               & IR shoulder of 122.7\cm in Pol.~1\\
\hline

\textbf{122.7}   &  1,2   & --               & Magnetic excitation \\
\hline

\textbf{128.5}   &  1     & 129.7 ($A_u$)    &  \\
\hline

\textbf{133.1}   &  1,2   & --               & Shows Faraday angle \\
\hline

136     &  1,2   & --               &  \\
\hline

141     &  1,2   & 139.0 ($A_u$)    & Shoulder of 146.7\cm in Pol.~1 \\
\hline

\textbf{146.7}   &  1     & --               & Strong IR mode in Pol.~1; $\delta_s\mathrm{PX}_3$ mode~\cite{Jouanne1988} \\ 

        &        &                  & Wide shoulder next to mode 141\cm in Pol.~2  \\
\hline

\textbf{153}     &  2     & --               &  \\
\hline

156     &  1,2   & 153.8 ($A_u$)    & Shoulder of 146.7\cm in Pol.~1 \\ 
        &        &                  & and 160.3 \cm in Pol.~2 \\
\hline

160.3   &  2     & 161.7 ($B_u$)    & Ref.~\citep{Jouanne1988} \\
\hline

163.5   &  1     & --               & Shoulder of 165.2\cm in Pol.~1 \\
\hline

165.2   &  1     & --               &  \\
\hline

165.6   &  2     & 167.6 ($B_u$)    & Shows Faraday angle \\
\hline

173     &  1     & --               & Shoulder of 184.6\cm in Pol.~1 \\ 
\hline

174.7   &  2     & 174.5 ($B_u$)    & \\ 
\hline
      
176.1   &  1     & 178.0 ($A_u$)    & \\
\hline

183.6   &  2     & 179.3 ($B_u$)    & \\
\hline

\textbf{184.7}   &  1     & --               & Strong IR mode; $\mathrm{R}'_{xy}\mathrm{PX}_3$ mode~\cite{Jouanne1988} \\
\hline

\textbf{189.3}  &  2     & --               & Strong IR mode in Pol.~2; $\mathrm{R}'_{xy}\mathrm{PX}_3$ mode~\cite{Jouanne1988} \\ 
        &        &                  & Shoulder of 184.7\cm in Pol.~1 \\
\hline
        
207     &  2     & 202.5 ($B_u$)    & \\
\hline

\textbf{215.4}   &  1,2   & 211.8 ($A_u$)    & IR mode in Pol.~2\\
        &        &                  & Weak mode in Pol.~1 \\
\hline

227.7   &  2     & 230.3 ($B_u$)    & \\
\hline

228.4   &  1     & 221.6 ($A_u$)    & \\
\hline

232.2   &  2     & 231.9 ($B_u$)    & \\
\hline

251     &  1,2   & --               & IR shoulder of 257.4\cm in Pol.~1 \\ 
        &        &                  & and 260.1\cm in Pol.~2\\
\hline
        
255.6   &  2     & --               & IR shoulder of 260.1\cm\\
\hline

\textbf{257.6}  &  1     & 254.6 ($A_u$)    & Strong IR mode; $\delta_d\mathrm{PX}_3$ mode~\cite{Jouanne1988} \\
\hline

\textbf{260.1}  &  2     & --               & Strong IR mode; $\delta_d\mathrm{PX}_3$ mode~\cite{Jouanne1988} \\
\hline

263.3   &  1     & --               & \\
\hline

268.8   &  2     & --               & \\
\hline

274.5   &  1     & --               & \\
\hline

306     &  1     & --               & \\
\hline

375.6   &  1     & --               & Wide IR feature \\
\hline

532     &  1     & 538.9 ($A_u$)    & \\
\hline

546.1   &  1     & 541.7 ($A_u$)    & \\
\hline

553.2   &  2     & 550.3 ($B_u$)    & \\
\hline

556.3   &  1     & 556.7 ($A_u$)    & \\
\hline

559.2   &  2     & --               & \\
\hline

561.4   &  1     & --               & \\
\hline

573.5   &  1     & --               & Strong IR mode \\
\hline

575     &  2     & --               & Strong IR mode \\

\hline
\end{tabular}
\label{tab:IRphononlist}
\end{table*}

\subsection{The 122\cm Excitation in FePS$_3$}

\begin{figure}
    \centering
    \includegraphics[width=\linewidth]{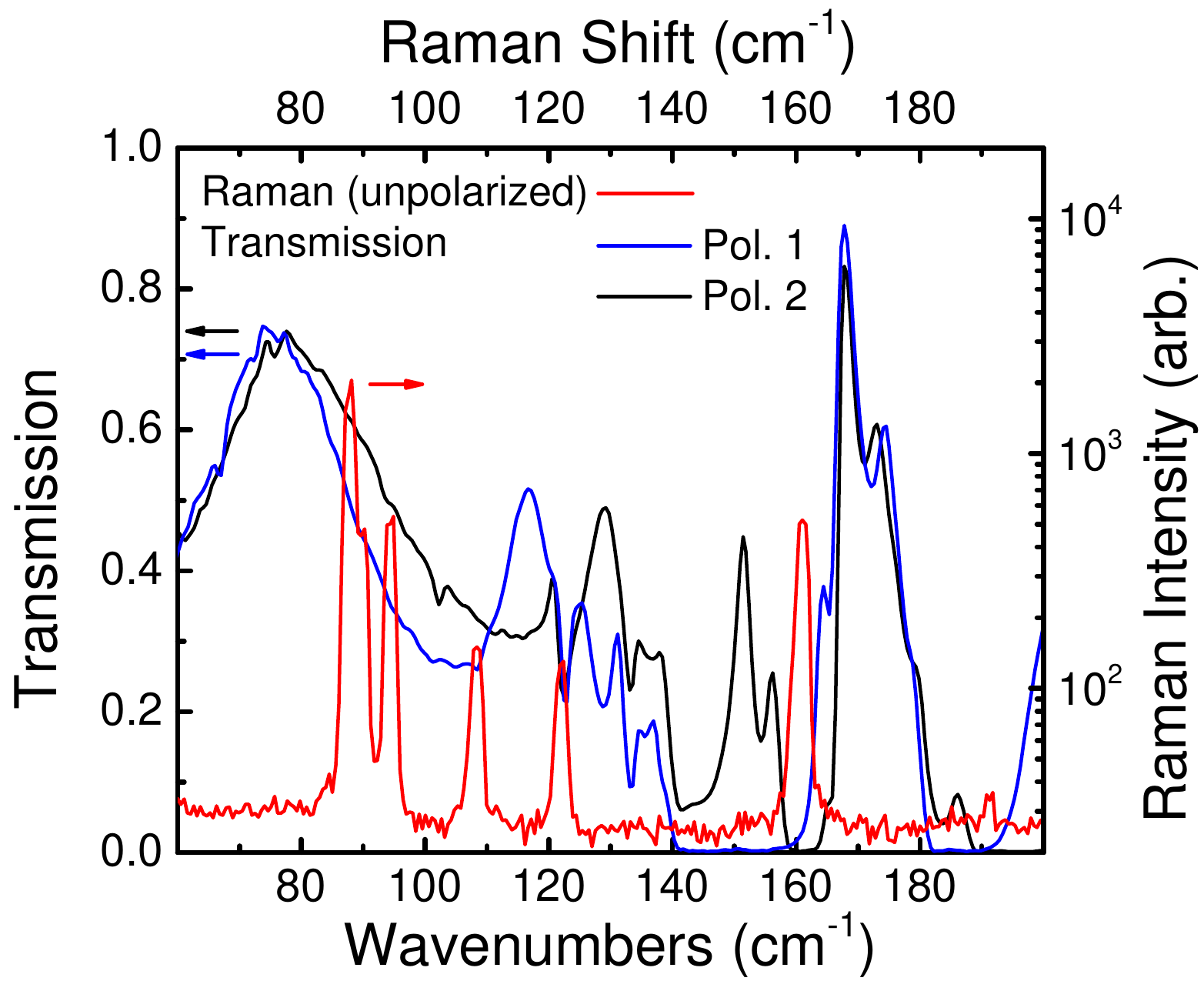}
    \caption{Comparison of the Raman microscopy spectrum (red) with the transmission spectra for the two principal optical polarizations (black and blue) measured at 5~K and zero magnetic field. The comparison shows the 122\cm excitation in all spectra (The negligible slight discrepancy in the Raman shift is acceptable within our systematic error and can be attributed to the calibration of the elastic line in the Raman spectrometer).}
    \label{fig:TrvsRaman}
\end{figure}

Despite the anisotropy seen in the phononic spectra, the mode at 122\cm, identified previously as the 122\cm magnon of FePS$_3$~\cite{Ghosh2021, Scagliotti1987,Lanon2016} and, as detailed in Appendix~\ref{sec:CFE}, is inherent to the Fe $3d^6$ multiplet structure, is present in both polarizations with the same spectral weight and at the same frequency, up to the spectral resolution of our measurements (about 0.5\cm). This is consistent with the 122~\cm\ excitation being isotropic in FePS$_3$ compared to the aforementioned lattice vibrational modes. Notably, the excitation at 122\cm appears in both the Raman shift and the infrared spectra as shown in Fig.~\ref{fig:TrvsRaman}. Although this excitation is observed in both IR and Raman spectroscopy, which might imply inversion symmetry breaking resulting in the mixing of odd and even modes, it is important to note that inversion symmetry is preserved in FePS$_3$ at the AFM state~\cite{Lanon2016,Chu2020,Ni2022}. We note that recent work has discussed related effects in FePS$_3$~\cite{Geraffy2026}, however, the interpretation of these observations remains an open question.

\begin{figure}
    \centering
    \includegraphics[width=1.0\linewidth]{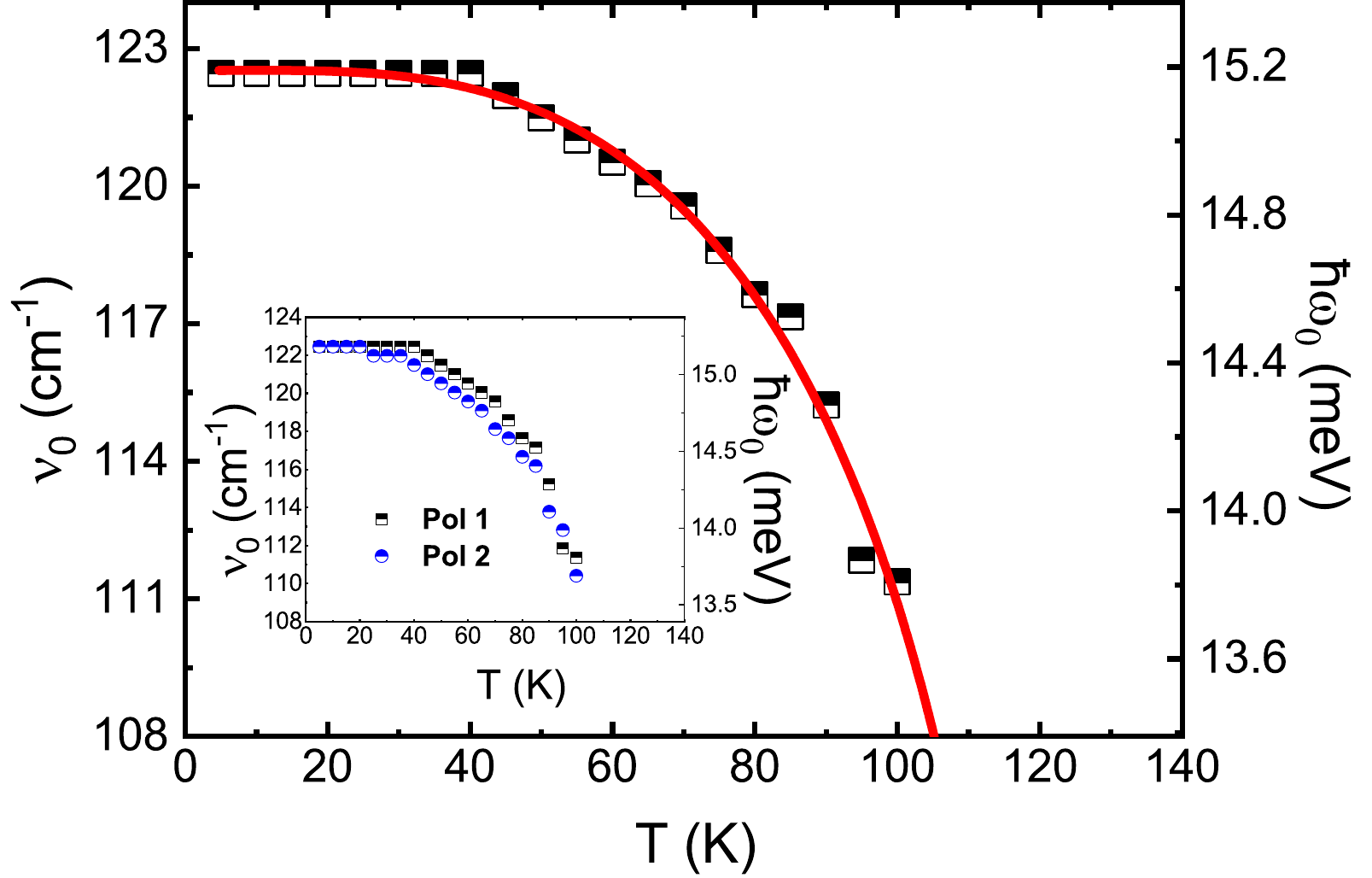}
    \caption{Central frequency of the 122\cm excitation as a function of reduced temperature ($T_N/T$) at zero magnetic field. The red line shows a fit to an order parameter temperature dependence (Eq.~\ref{Eq:OPfit}). The fitting yields $a = 33.39 \pm 0.87$, $c = 89.13 \pm 0.80$ (Pol 1-the solid red line) and $a = 33.94 \pm 1.01$, $c = 87.98 \pm 0.94$ (Pol 2). The inset compares both principal polarizations, indicating isotropic temperature dependence.}
    \label{fig:OmegaT}
\end{figure}

We will now turn to the temperature- and magnetic-field-dependent 122\cm excitation. The mode hardens as the temperature is lowered and follows the same temperature dependence for both principal polarizations. Figure~\ref{fig:OmegaT} shows the mode's central frequency as obtained by the Drude-Lorenz fit to our transmission data for both principal axes. Since the temperature dependence resembles that of an order parameter change, we employed the following expression to fit our data: 
\begin{equation}
   \omega_0 (T) = a \cdot \tanh\left[1.82\left(1.018\left(\frac{T_{N}}{T} - 1\right)\right)^{0.5}\right] + c
    \label{Eq:OPfit}
\end{equation}
where $a$ and $c$ are free parameters of the fit (see caption of Fig.~\ref{fig:OmegaT} for values) and $T_N=118~K$ (See Eq.~5 in Ref.~\citep{Carrington2003} and relevant examples in Ref.~\citep{Senapati2011,Park2024}). The mode's central frequency $\omega_0$ follows the temperature dependence of a magnetic order parameter with transition temperature $T_N$ and provides clear evidence that the 122\cm excitation originates at the antiferromagnetic state.   

\begin{figure}
    \centering
    \includegraphics[width=\linewidth]{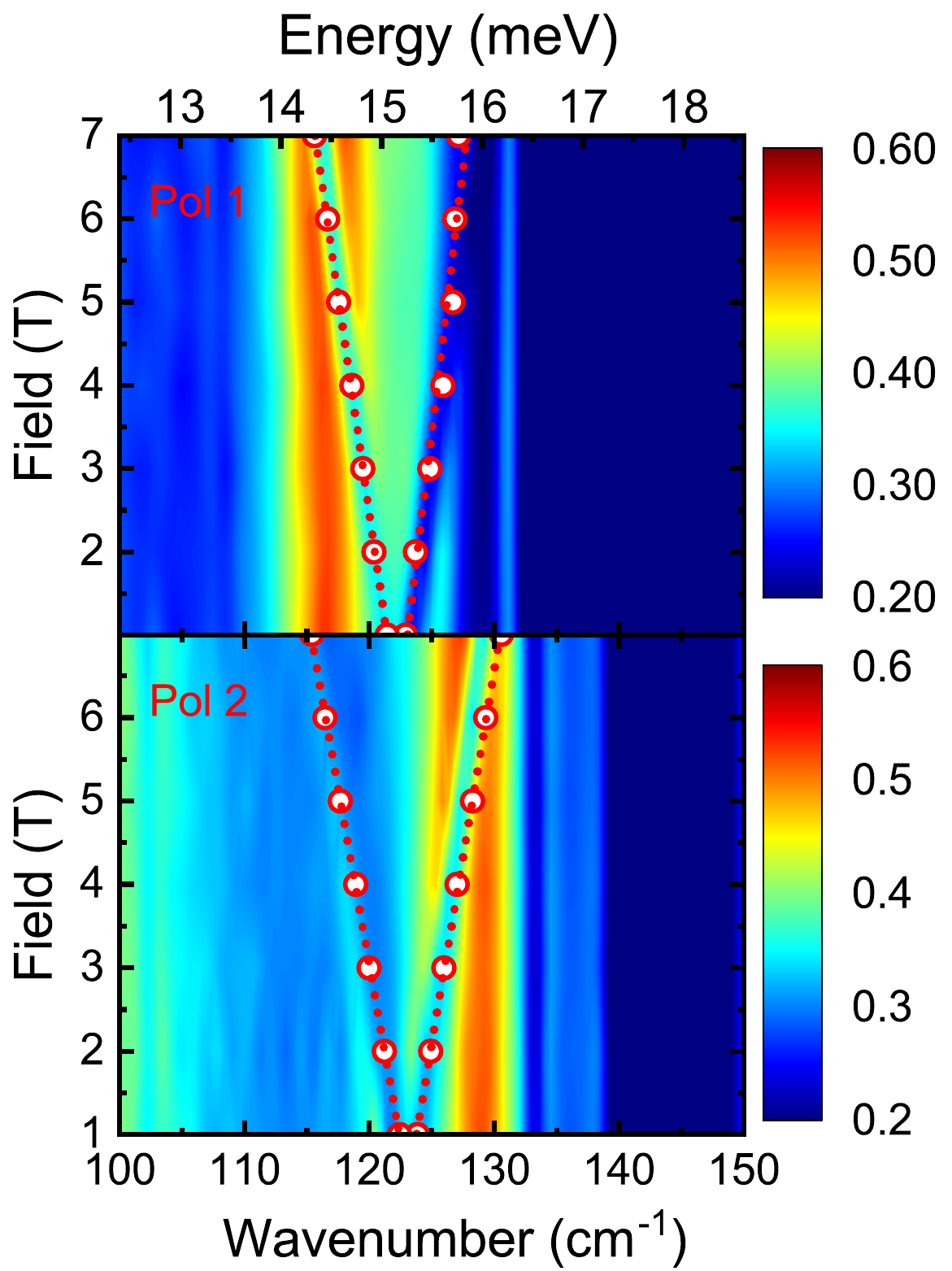}
     \caption{Polarized magneto-transmission color maps of FePS$_3$ at 5~K for two principal polarizations, measured as a function of magnetic field (up to 7~T in 1~T steps). Circles mark the central frequencies of the split 122\cm mode, highlighting the field-induced splitting behavior.}
    \label{fig:Field_Pol}
\end{figure}

\begin{figure}
    \centering
    \includegraphics[width=\linewidth]{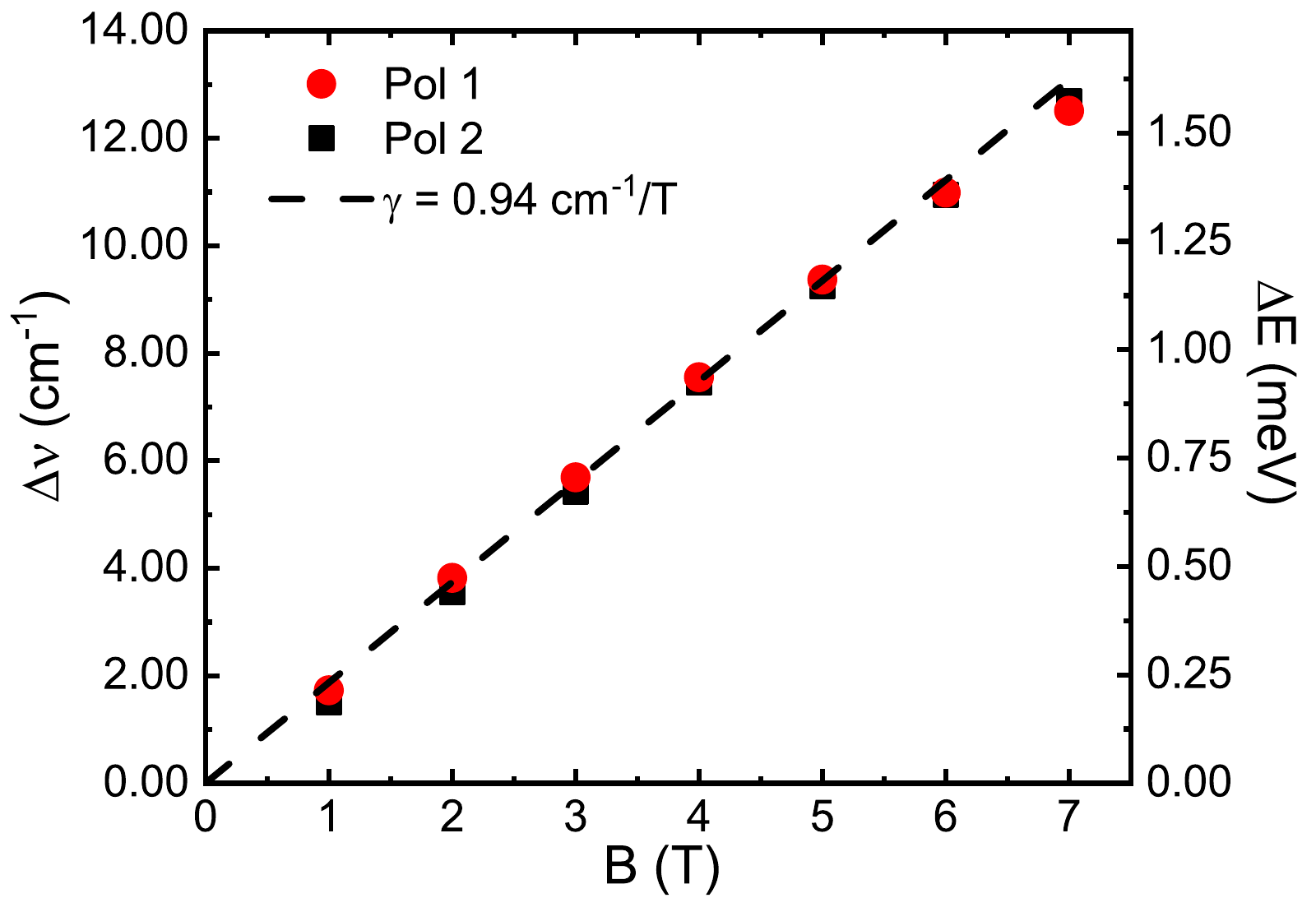}
    \caption{Magnetic-field dependence of the energy splitting \(\Delta \nu\) between the central frequencies of the 122~cm\(^{-1}\) mode in FePS$_3$, measured for the two principal polarizations (Pol 1 and Pol 2) at 5~K. The splitting \(\Delta \nu\) is extracted from fits to the transmission spectra and plotted as a function of magnetic field \(B\). The dashed line represents the expected linear dependence assuming a gyromagnetic ratio (\(\gamma \approx 0.94\)~cm\(^{-1}\)/T), which holds up to \(\sim 7\)~T before deviations arise due to collective modes' interactions~\cite{Vaclavkova2021}.}
    \label{fig:Omega_B}
\end{figure}

Another compelling piece of evidence for the magnetic origin of the 122\cm excitation was shown in recent Raman spectroscopy measurements of FePS$_3$~\cite{McCreary2020} where the excitation, previously considered as a Brillouin-zone folded phonon branch~\cite{Balkanski1987,Zhang2021}, splits into two modes as a function of the magnetic field. Figure~\ref{fig:Field_Pol} shows the field-induced splitting of the 122\cm mode in our transmission spectra, consistent with similar observations in recent optical spectroscopy studies~\cite{McCreary2020,Vaclavkova2021}. This splitting becomes clearly resolved at fields up to 7~T. The mode's splitting follows a field dependence with a gyromagnetic ratio of a free electron, and as shown in Figure~\ref{fig:Omega_B} for our data in the IR spectra, consistent with previous Raman measurements~\cite{McCreary2020}. However, this field dependence breaks down at higher magnetic fields due to the avoided crossing of two modes, the lower energy mode's branch and the Raman mode at 108\cm~\cite{Vaclavkova2021}. We will elaborate on another aspect of a spin-phonon branch avoided crossing, however, for the upper mode.

\subsection{Faraday Angle Rotation}

\begin{figure*}
    \centering
    \includegraphics[width=0.8\linewidth]{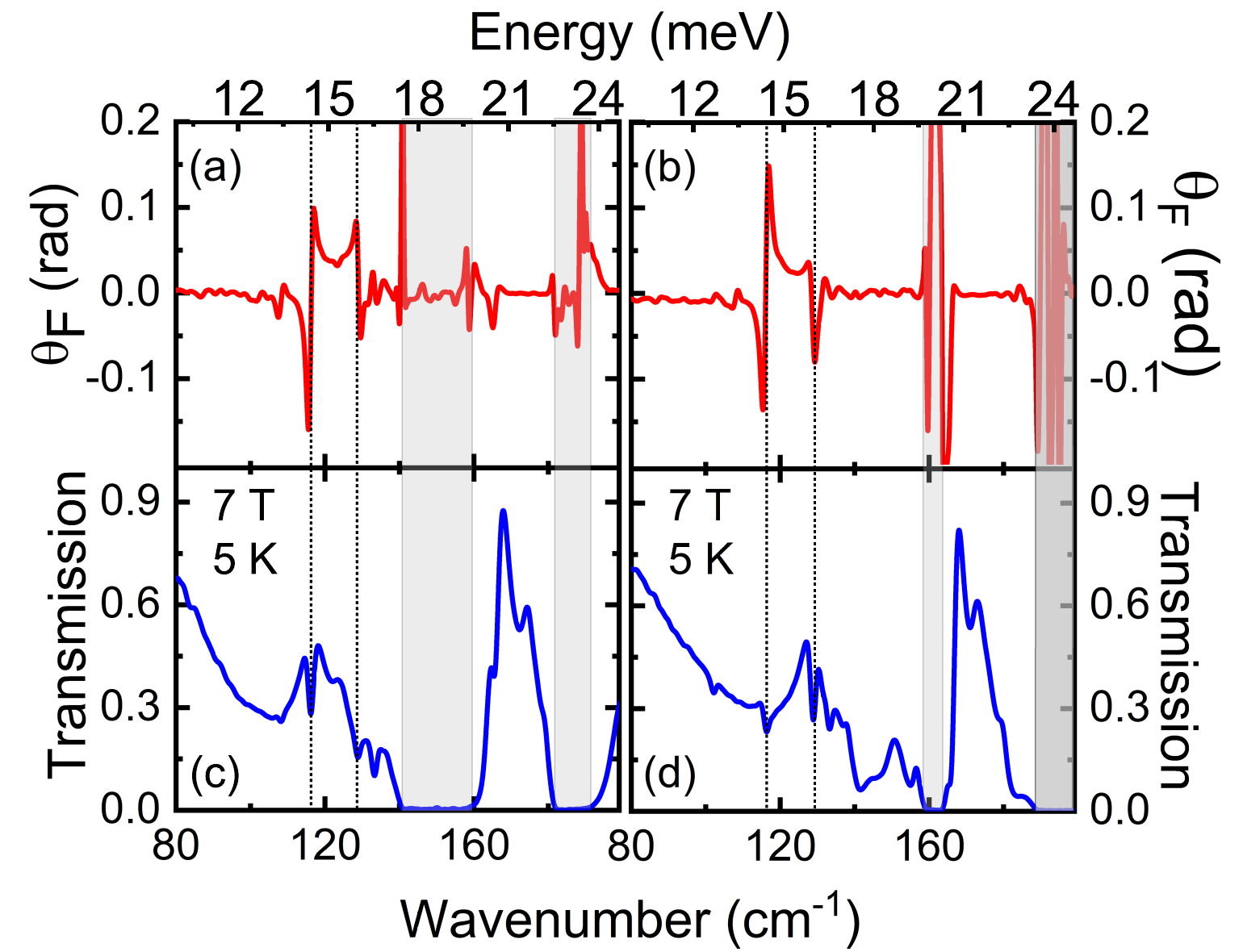}
  \caption{Faraday rotation angle (top panels) and corresponding magneto-transmission spectra (bottom panels) of FePS$_3$ measured at 5~K under $\pm7$~T magnetic field for two FIR polarizer orientations: (a,c) Pol~1 and (b,d) Pol~2. Pronounced Faraday rotation features appear at the 122\cm excitation, showing symmetric splitting with field, as well as at several phonon modes, highlighting strong spin–phonon coupling in the antiferromagnetic state. Transmission below 1\% is masked (gray) to avoid unreliable Faraday angle extraction.}
    \label{fig:Faraday}
\end{figure*}

In addition to the infra-red spectroscopy of the temperature and magnetic field dependence of the 122\cm excitation, we show in Figure~\ref{fig:Faraday} the Faraday angle rotation at 5~K and fields of $B=\pm 7~T$, which was not reported before for FePS$_3$, to the best of our knowledge. The measurement was obtained using the full protocol for the two principal axes used in this work, as explained in the previous section [Sec. \ref{sec:optics}].  

The Faraday angle rotation is shown in Fig.~\ref{fig:Faraday} along with the transmission as measured by the parallel polarization configuration of the analyzer and polarizer. To avoid artifacts in regions with very low signal, we masked out spectral regions where the transmission coefficient dropped below 1\%. This threshold lies above the sensitivity limit of our experimental setup, ensuring the reliability of the Faraday angle measurements. In this way, we can discuss several interesting features, starting with the collective excitations that originate in the antiferromagnetic state and then with the lattice collective modes that are coupled to the magnetic order. 

The 122\cm excitation that splits into two modes shows a Faraday angle rotation of about 0.1 to 0.15~rad at 5~K and 7~T. This Faraday angle is much larger than the sensitivity of the measurements setup (about 2~mRad) and well above the measurement noise level, as shown in Figure~\ref{fig:Faraday}. 

\begin{figure}
    \centering
    \includegraphics[width=\linewidth]{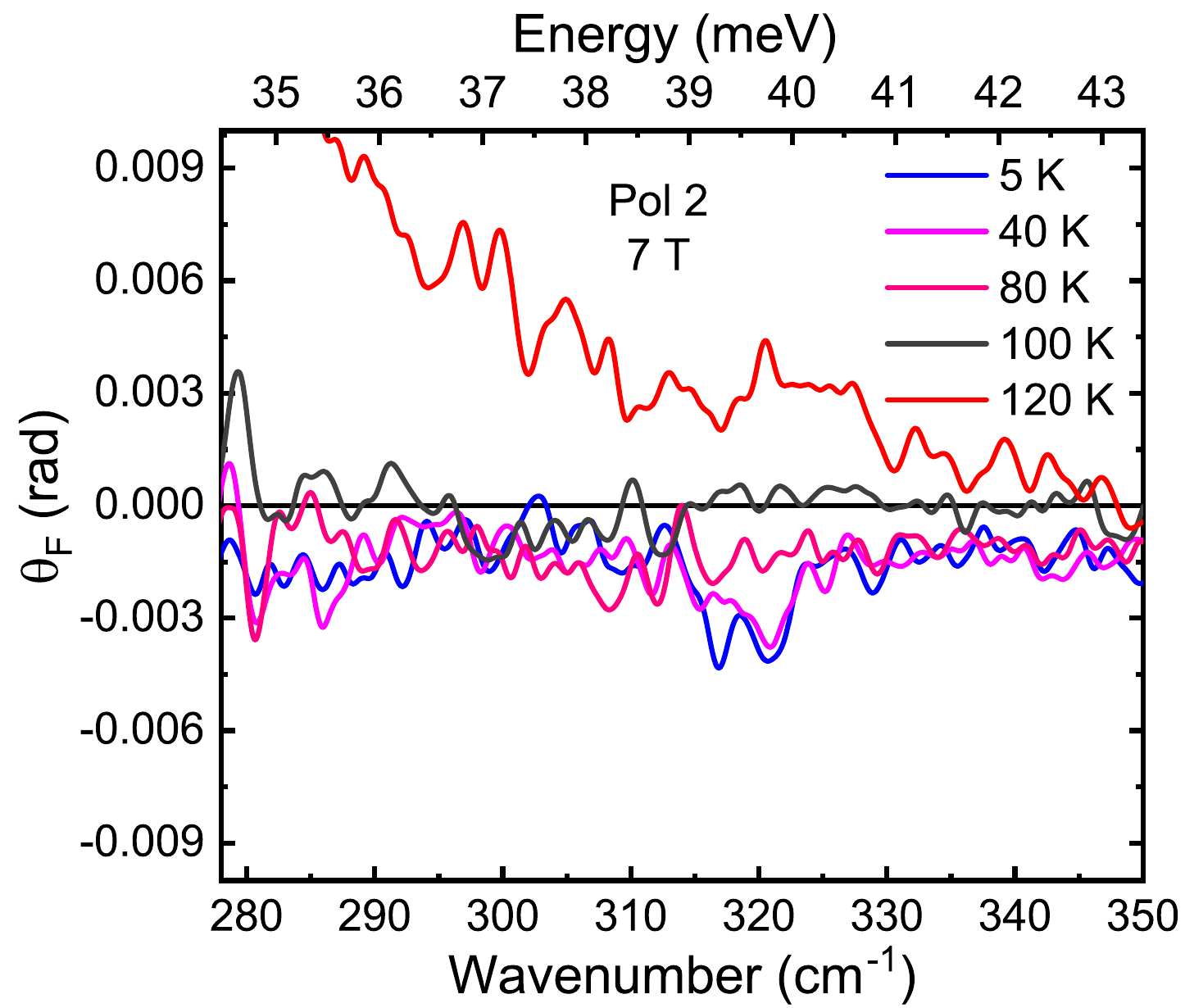}
  \caption{Temperature evolution of the Faraday rotation angle in the 280–350\cm range of FePS$_3$, measured at 7~T with FIR polarizer orientation Pol~2. A prominent feature near 320\cm emerges below the N\'{e}el temperature (T$_N$ = 118~K), increases in strength with decreasing temperature, and is attributed to a higher-energy mode at 40~meV. This temperature-dependent behavior further supports the magnetic origin of the excitation.}

    \label{fig:Faraday_320}
\end{figure}

In addition to the 122\cm, it is known that FePS$_3$ has another magnetic excitation at higher energies of about 40~meV~\cite{Lanon2016}. A close examination of our Faraday angle rotation spectrum near that energy reveals a very weak feature, which we attribute to this high-energy magnetic excitation. Figure~\ref{fig:Faraday_320} shows the Faraday rotation angle for selected temperatures above and below the antiferromagnetic transition of FePS$_3$. Below $T_N$ but still close to the phase transition, it is very hard to detect any Faraday angle rotation above our measurement baseline and within the noise level; however, as the system goes deep into the antiferromagnetic phase, a dip in the data appears at about 320\cm, which corresponds to an energy of about 40~meV. The appearance of a Faraday rotation signal at 320\cm, accompanied by a temperature-dependent shift and an intensity increase, provides evidence that this feature originates from a magnetic excitation.   

It is important to note here that the frequency 122\cm in our data corresponds to the energy of 15~meV detected at the commensurate $q=(0,1,0)$ momentum vector via inelastic neutron scattering measurements~\cite{Lanon2016} with a momentum dispersion reaching about 17~meV at $q=(0,0.5,0)$. On the other hand, the excitation at about 40~meV is seen at the incommensurate momentum of, e.g., $q=(\pm1,k,\mp1/3)$ with a momentum dispersion reaching about 17~meV at e.g., $q=(\pm2,k,\mp2/3)$ (where $k$ is $0.5,1,1.5$). In general, optical spectroscopy is sensitive only to $q\approx0$ collective excitations. Nevertheless, the frequencies at which we measure the Faraday angle rotation signal agree with the excitation energies of FePS$_3$ as measured by inelastic neutron scattering. Moreover, the difference in intensity between the excitations at 122~\cm\ (15~meV) and 320~\cm\ (40~meV) is observed experimentally; however, its microscopic origin remains an open question and requires further investigation.    

\section{Discussion\label{sec:discussion}}

\subsection{Crystal Symmetries in FePS$_3$}

FePS$_3$ has a monoclinic $C2/m$ structure and $C_{2h}$ space group in the paramagnetic state, with a twofold rotation axis along $b$ and a mirror plane in the $ac$ plane. The structure also preserves inversion symmetry~\cite{Lanon2016,Chu2020,Ni2022}, and no symmetry lowering is assumed in the present analysis. Although the lattice is intrinsically anisotropic, the transmission spectra appear nearly isotropic above $T_N$ (e.g., at $T=150$~K) for both polarizations, with only small differences in phonon energies of up to several \cm, particularly for modes near 145, 185, 255, and 570~\cm.

In this symmetry, the infrared-active phonons belong to the $A_u$ and $B_u$ irreducible representations, corresponding to dipole moments parallel and perpendicular to the crystallographic $b$-axis, respectively. Accordingly, polarization-resolved infrared spectroscopy directly probes the directional character of the lattice dynamics. The $A_u$ modes are associated with displacements along the $b$-axis, while the $B_u$ modes involve motion within the $ac$ plane, where a larger number of deformation and librational modes of the P$_2$S$_6$ units are expected at low energies. Comparing the polarized spectra with the DFT phonon calculations indicates that Pol.~2 is predominantly associated with $B_u$ modes and therefore with the $a$-axis, while Pol.~1 corresponds mainly to $A_u$ modes and the $b$-axis. This assignment is supported by the richer low-energy phonon structure observed in Pol.~2 below 250~cm$^{-1}$, consistent with the in-plane character of the $B_u$ vibrations.

Several calculations of the phonon spectrum of FePS$_3$ have been reported in the literature. Bernasconi~\textit{et al.}~\cite{Bernasconi1988} predicted $A_u$ and $B_u$ doublets near 150, 221, 258, and 570~\cm. In our calculations, the corresponding modes appear at slightly different frequencies, as summarized in Table~\ref{tab:IRphononlist}. For example, the mode near 570~\cm corresponds to calculated modes at 550~\cm ($B_u$) and 556~\cm ($A_u$), while the modes near 258~\cm and 150~\cm correspond to calculated modes near 254~\cm and 154~\cm, respectively. In addition, our calculations yield modes at 178~\cm ($A_u$) and 179~\cm ($B_u$), which are closer to the experimentally observed features than the previously suggested 221~\cm mode. Despite these differences, the assignments in Table~\ref{tab:IRphononlist} are based on the closest correspondence between calculated and experimental frequencies, independent of spectral weight.

Below $T_N$, the zigzag antiferromagnetic order lifts the in-plane equivalence between the crystallographic $a$ and $b$ directions, resulting in a pronounced anisotropy in the phonon response. This anisotropy is particularly evident in the low-energy region, where several modes exhibit polarization-dependent frequency shifts and spectral weights. This behavior reflects spin--phonon coupling in the magnetic ground state: lattice vibrations that modulate the Fe--S--Fe exchange pathways experience different effective force constants depending on their polarization. In particular, low-energy deformation and librational modes of the P$_2$S$_6$ units are expected to couple strongly to these exchange interactions.

In addition to anisotropic spin--phonon coupling, the magnetic ordering in FePS$_3$ leads to an enlargement of the magnetic unit cell, both within the $ab$ plane and along the $c$-axis. As a result, phonon modes with finite momentum in the paramagnetic phase can be folded back to the Brillouin-zone center in the antiferromagnetic state. Such Brillouin-zone-folded phonons have been reported in previous Raman spectroscopy studies~\cite{Wang2016,Lee2016,Hashemi2017,McCreary2020}, where additional modes appear below $T_N$, as we also confirmed in our supplementary Raman spectroscopy measurements. In our polarized infrared data, some of the weak spectral features observed in the low-energy region may similarly originate from folded phonons activated by the magnetic superstructure. While a detailed assignment of these modes is beyond the scope of the present work, their appearance together with the pronounced polarization dependence is consistent with contributions from both Brillouin-zone folding and anisotropic spin--phonon coupling, with the latter providing the dominant contribution to the modified phonon spectrum in the antiferromagnetic phase and motivating further theoretical work.

To further probe this coupling, we compare the polarized infrared transmission data with Raman spectroscopy performed on crystals from the same growth batch (Fig.~\ref{fig:TrvsRaman}). Several vibrational modes exhibit a measurable Faraday rotation, indicating that lattice excitations acquire a magnetic-field-dependent optical response.

In particular, the Raman-active mode near 108\cm appears as a weak feature in the infrared spectra and develops a finite Faraday rotation of approximately 25~mrad. At high magnetic fields, an avoided crossing is observed between this mode and the lower-energy branch of the split 122\cm excitation~\citep{Wyzula2022}. In addition, infrared-active modes at 133.1~\cm and 165.6~\cm exhibit significant Faraday rotation, reaching values of about 25~mrad and 50~mrad, respectively, at 7~T (Fig.~\ref{fig:Faraday}).

The polarization dependence of the Faraday rotation is particularly pronounced in the low-energy region, consistent with the anisotropic exchange interactions of the zigzag antiferromagnetic structure. Phonon modes associated with the $a$- and $b$-axes therefore experience distinct renormalization, providing further evidence that spin--phonon coupling in FePS$_3$ is intrinsically anisotropic.

\begin{figure}
    \centering
    \includegraphics[width=\linewidth]{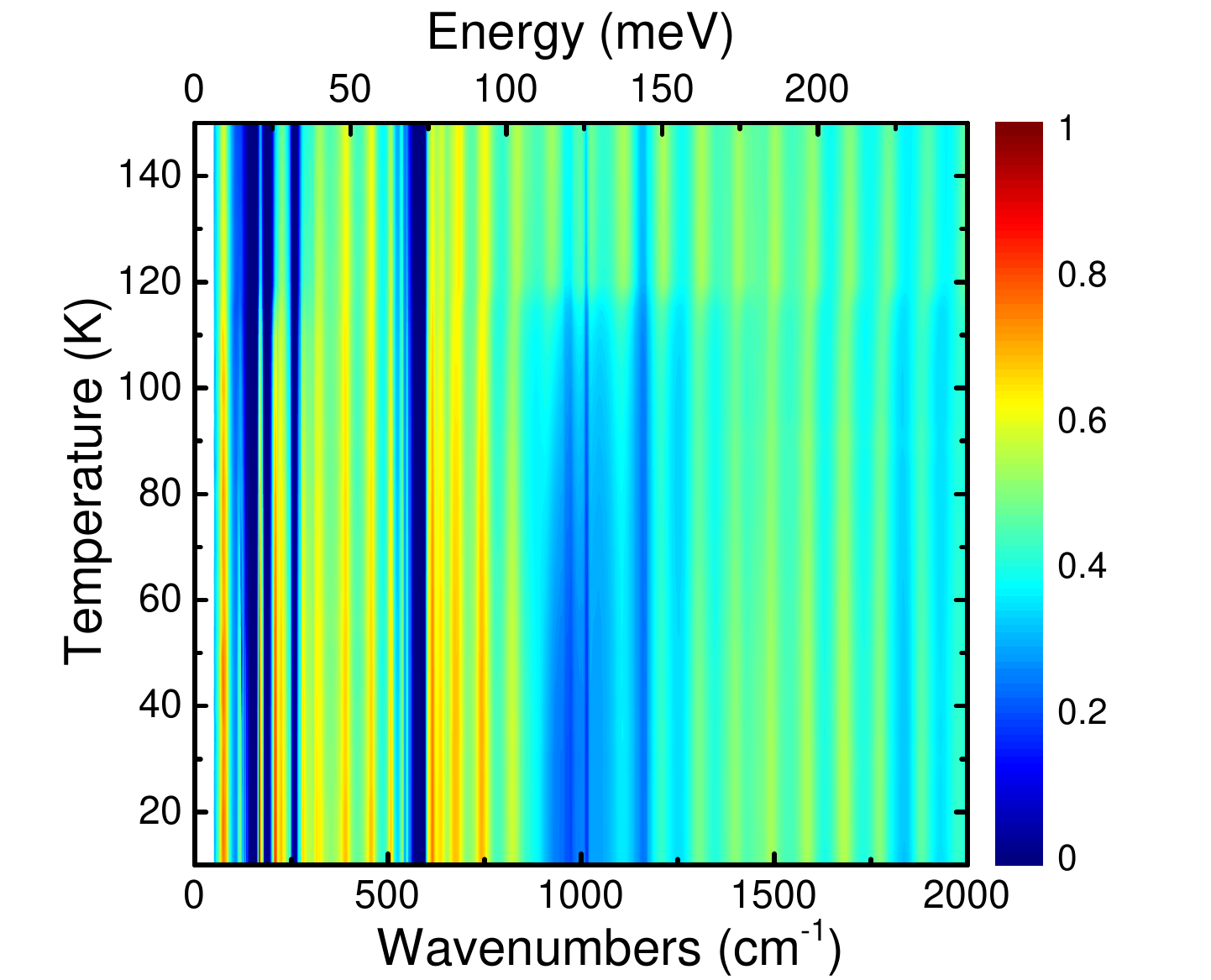}
    \includegraphics[width=1.01\linewidth]{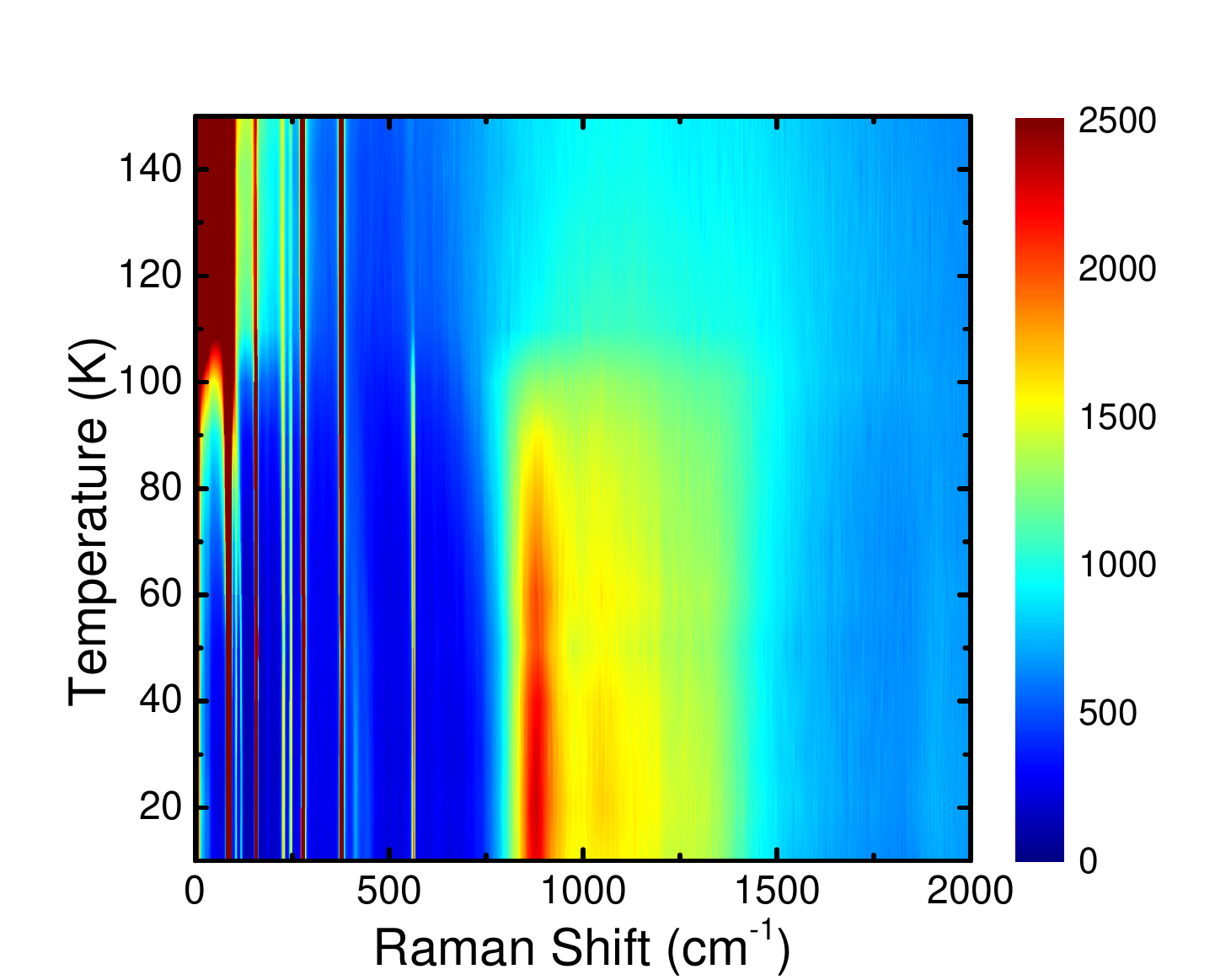}
    \caption{Top panel: Color map of polarized infrared transmission (from 50\cm and for Pol 1) at zero field as a function of temperature. Bottom panel: 2D color map of the unpolarized Raman Intensity in zero field as a function of temperature. The temperature range in both measurements was obtained between 10~K to 150~K in intervals of 10~K.}
    \label{fig:MIR_Trans}
\end{figure}

At higher energies, a broad feature centered around 900~\cm is observed in the mid-infrared transmission (Fig.~\ref{fig:MIR_Trans}), primarily in the Pol.~2 channel and below $T_N$. A corresponding feature is also present in the Raman spectra. While its origin remains unclear, it may be associated with multi-phonon excitations, which are known to produce broad spectral features at higher energies. In particular, combinations of optical phonons or overtones can give rise to continua in this spectral range. The appearance of this feature in both spectroscopies and its sensitivity to the magnetic phase may be consistent with a coupling between lattice and electronic or magnetic degrees of freedom (see Appendix~\ref{sec:CFE}).

Overall, our results show that polarization-resolved magneto-optical spectroscopy provides direct access to anisotropic spin--phonon coupling in FePS$_3$, linking the lattice dynamics to the underlying zigzag antiferromagnetic order.

\subsection{Dichroism of the Optical Circular Conductivity}

FePS$_3$ is considered to have an Ising-like (1D) spin Hamiltonian where the spins are directed along the $c$-axis. Hence, there should be a negligible, if any, anisotropy effect on the absorption spectra in different polarizations along the crystal's $ab$ plane, as can be seen in our data (see App.~\ref{sec:appTr}). This allows us to consider the 122\cm as an isotropic excitation in the polarized magneto-transmission data. Since we can quantitatively obtain the absolute transmission, the complex conductivity, and the Faraday angle of FePS$_3$ in our measurements, it is, therefore, possible to extract from our data the circular optical conductivity of the split 122\cm mode. The two field-separated collective modes can be associated with the right- and left-handed circular polarizations of the two opposite spin sublattices in the antiferromagnetic state ~\cite{Lan2017}. At zero field, there is no difference between one spin population and the other, and the 122\cm excitation has a single eigen energy. Hence, the two circular polarizations combine into one linearly polarized mode at 122\cm. The application of a magnetic field splits the modes into low and high energy modes with a corresponding circular optical conductivity $\sigma_+$ and $\sigma_-$.

We follow the procedure described in Ref.~\citep{Levallois2015} however, due to the anisotropy in the magnetic state, the in-plane optical conductivity tensor should be described as follows:
\begin{equation}
    \sigma = 
    \begin{pmatrix}
    \sigma_{xx} & \sigma_{xy} \\
    -\sigma_{xy} & \sigma_{yy}
    \end{pmatrix}
\end{equation}
where the in-plane anisotropy of the FePS$_3$ optical conductivity is given by different dynamic conductivity functions $\sigma_{xx}$ and $\sigma_{yy}$. The latter terms capture the ordinary absorption and the birefringence of the sample, while the off-diagonal term, $\sigma_{xy}$, is associated with the magneto-optical effects and the optical Hall conductivity. We note that FePS$_3$ has monoclinic symmetry, for which the dielectric tensor generally contains additional off-diagonal elements beyond the form written above. However, for our experimental geometry, where optical propagation is along the $c$-axis and the measured response is confined to the $ab$ plane, these contributions are considered small within our experimental configuration. In this sense, the monoclinic distortion is retained, but its effect on the in-plane optical response is weak, allowing us to adopt an effective orthorhombic approximation. This approximation captures the dominant in-plane response relevant for the present analysis.

Plugging in the eigenvectors of the Right- and Left-handed circular polarized light $e_{\pm}=1/\sqrt{2}\left(1,\pm i\right)$ results in the circular optical conductivity:
\begin{equation}
    \sigma_{\pm}(\omega)=\frac{\sigma_{xx}(\omega)+\sigma_{yy}(\omega)}{2} \pm i \sigma_{xy}(\omega)
\label{Eq:sigma_pm}
\end{equation}
for the Right-handed ($\sigma_{+}$) and Left-handed ($\sigma_{-}$) optical circular conductivity. We note that $\sigma_{xy}$ is expected to be only weakly dependent on the specific in-plane axis used in the analysis within this approximation. Moreover, while linear birefringence mixes the two circular eigenmodes, so they are not perfectly circular, it does not affect the Faraday angle rotation since it is not related to the off-diagonal terms and cancels out in the cross-polarization measurement. However, the values of $\sigma_{xy}$ can be different for the measurement and analysis of the Faraday angle rotation in Pol1 vs Pol2 due to systematic errors, e.g. residual cross polarization signal, and/or crystal misalignment. Our results show that throughout most of the spectral range, including the lower energy mode (approx. 116\cm) and phonons, there is a perfect agreement between the two sets of data with respect to $\sigma_{xy}$ obtained from the two in-plane polarizations. Some discrepancy between the values can be detected around the upper energy mode (approx. 129\cm, see App.~\ref{sec:appsigma}), which shows that it is local in frequency and probably not related to systematic errors or any sample tilt. Therefore, we have averaged the two $\sigma_{xy}$ outputs in order to obtain the optical circular conductivity $\sigma_{\pm}$.      

\begin{figure}
    \centering
    \includegraphics[width=\linewidth]{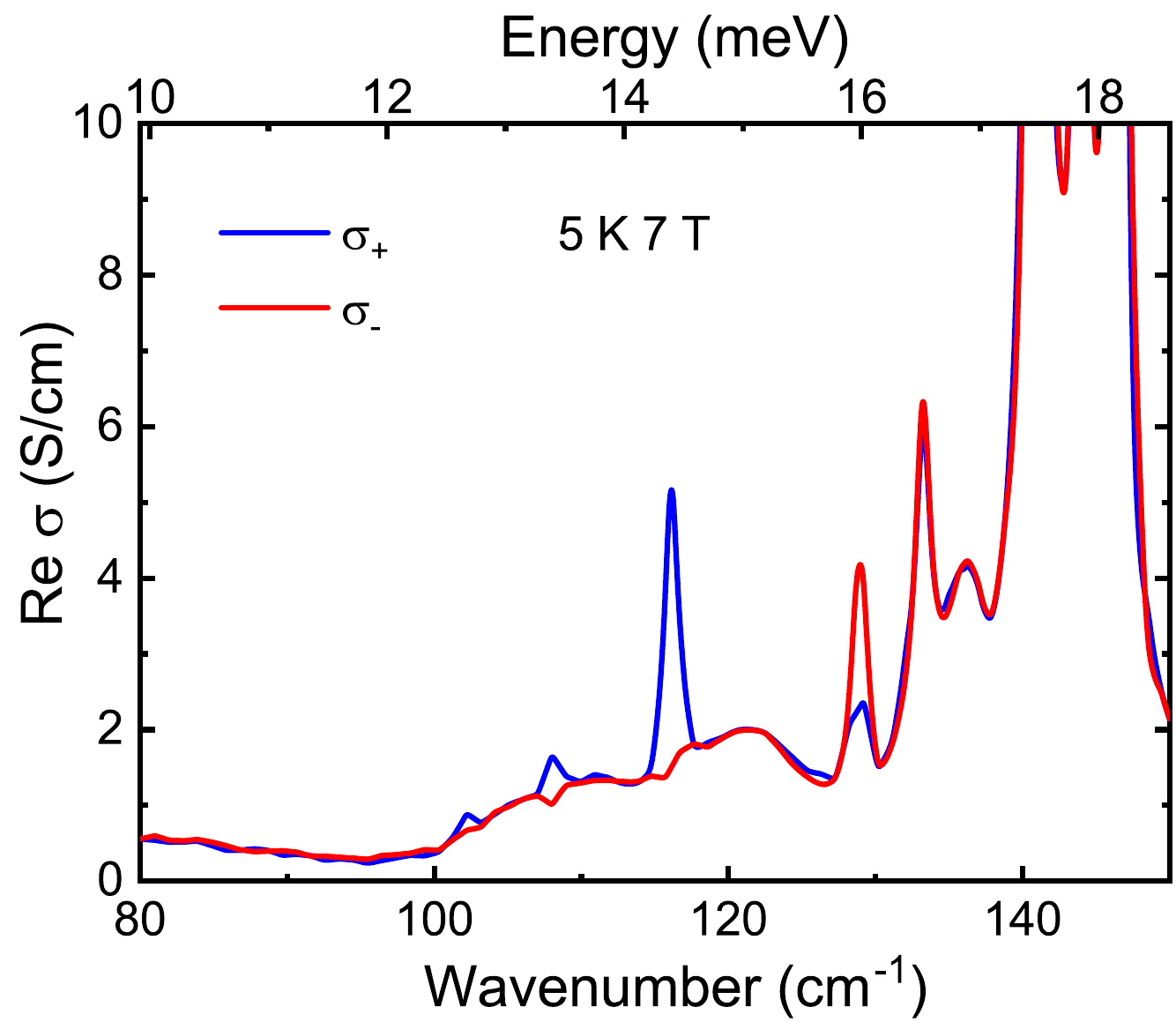}
    \caption{Circular optical conductivity $\sigma_{+}$ and $\sigma_{-}$ of FePS$_3$ at 5~K and 7~T, obtained by averaging the off-diagonal conductivity $\sigma_{xy}$ extracted independently for Pol 1 and Pol 2 using Eq.~\ref{Eq:sigma_pm}.}
    \label{fig:sigma_pm}
\end{figure}

Figure~\ref{fig:sigma_pm} shows the circular optical conductivity obtained from the analysis of the Magneto-Optical Kerr/Faraday Effect~\cite{Kuzmenko2005} and by taking into account the in-plane optical conductivity tensor as described above. As expected from a collinear antiferromagnetic resonance, there is a clear dichroism between $\sigma_{+}$ and $\sigma_{-}$ under the application of a magnetic field seen at frequencies of about 116\cm and 129\cm for 7~T for the lower energy and upper energy, respectively, of the 122\cm zero-field mode. 

In fact, Fig.~~\ref{fig:sigma_pm} shows that the lower mode is completely dichroic since the spectral weight of $\sigma_{-}$ in that excitation is the same magnitude as that of $\sigma_1$ at zero field while $\sigma_{+}$ shows a clear peak. On the other hand, the upper energy mode at around 129\cm does not show an ideal dichroic response with $\sigma_{+}$ being about factor of 1/3 from the value of $\sigma_{-}$. This response is not subject to the average of $\sigma_{xy}$ of the two polarizations and can be seen in both sets of data for the upper energy mode at 129\cm (See Appendix). 

Since this anomalous dichroism appears only in the upper energy mode (129\cm), which is also close to an infrared-active phonons at 128\cm and 133\cm, this behavior is consistent with spin--phonon coupling affecting the optical response and leading to a residual, yet significant, spectral weight in $\sigma_+$. Previous reports have identified coupling between the lower energy mode (116~\cm) and a Raman-active phonon at 108~\cm as an avoided crossing in the optical spectra~\cite{Wyzula2022}. In this context, our results may be consistent with an avoided transition involving the upper-energy mode (129~\cm) and the infrared-active phonon at 133~\cm already present at low fields.

As discussed above, the circular optical conductivity $\sigma_{\pm}$ reveals additional excitations beyond the field-split 122~\cm\ mode that exhibit a finite dichroic response. In particular, the mode near 108~\cm, which is primarily observed in Raman spectroscopy, also appears as a weak feature in the infrared data and develops a measurable Faraday rotation within our experimental sensitivity. We note that we do not attribute this observation to symmetry breaking, and its origin remains unclear. Possible contributions may arise from magnetic-dipole activity or from coupling to nearby infrared-active excitations. This behavior is consistent with the avoided crossing reported in high-field measurements~\cite{Wyzula2022}, and supports the presence of spin--phonon coupling in the collective excitation spectrum of FePS$_3$. We note that recent studies have shown that lattice vibrations in layered antiferromagnets can acquire finite angular momentum in the form of chiral phonons through magnon--phonon coupling, leading to circularly polarized phonon modes and magneto-optical activity~\cite{Cui2023}. While a detailed microscopic analysis is beyond the scope of the present work, such a mechanism may provide a qualitative framework for understanding the finite Faraday rotation observed in several phonon modes in FePS$_3$.

%Another Raman active mode at 160\cm can be detected in the circular optical conductivity in both polarizations due to a possible inversion symmetry breaking and leakage of the even vibrational modes into the odd modes IR spectra. Since the 160\cm mode is close to one of the IR active modes in our measurement, the Faraday angle of this mode is seen better in the second principal axis polarization shown in the right panel of Fig.~\ref{fig:Faraday}. In both cases, our data allow us to extract the spectral weight of the circular conductivity of these modes.

\section{Conclusions}

We reported a comprehensive polarization-resolved magneto-optical study of FePS$_3$, providing new insight into the interplay between lattice vibrations and magnetic order. In the antiferromagnetic phase, the phonon spectrum develops a pronounced in-plane anisotropy, reflecting the underlying zigzag magnetic structure and the associated directional exchange interactions. In contrast, the 122~\cm ($\sim15$~meV) excitation remains polarization-independent, exhibits an order-parameter-like temperature dependence, and splits under applied magnetic field, giving rise to a sizable Faraday rotation.

By combining absolute transmission measurements with quantitative Faraday rotation analysis, we extract the circular optical conductivities and reveal a clear dichroism of the split modes. The reduced dichroic response of the upper branch near 129~\cm is attributed to hybridization with a nearby infrared-active phonon, providing direct evidence for spin--phonon coupling. Additional phonon modes also exhibit finite Faraday rotation, indicating that lattice excitations acquire a magnetic-field-dependent optical response through their coupling to the spin system.

Our polarization-resolved measurements further show that the low-energy phonon spectrum is governed by both anisotropic spin--phonon coupling and the magnetic superstructure, which may activate additional folded phonon modes in the antiferromagnetic phase. These results establish polarization-resolved infrared spectroscopy combined with magneto-optical probes as a powerful approach for quantifying spin--lattice interactions and circular dichroism in low-dimensional antiferromagnets.

\section*{Acknowledgments}

This research was supported by the Israel Science Foundation (ISF) through project number 666/23. N.B. would like to acknowledge useful discussions with J. Teyssier, G. Blumberg, and M. Orlita. We thank Mustafa (Kasu) Bohra for his help with the crystal structure figure.    

\appendix 

\section{Transmission Data}\label{sec:appTr} 

\begin{figure}
    \centering
    \includegraphics[width=\linewidth]{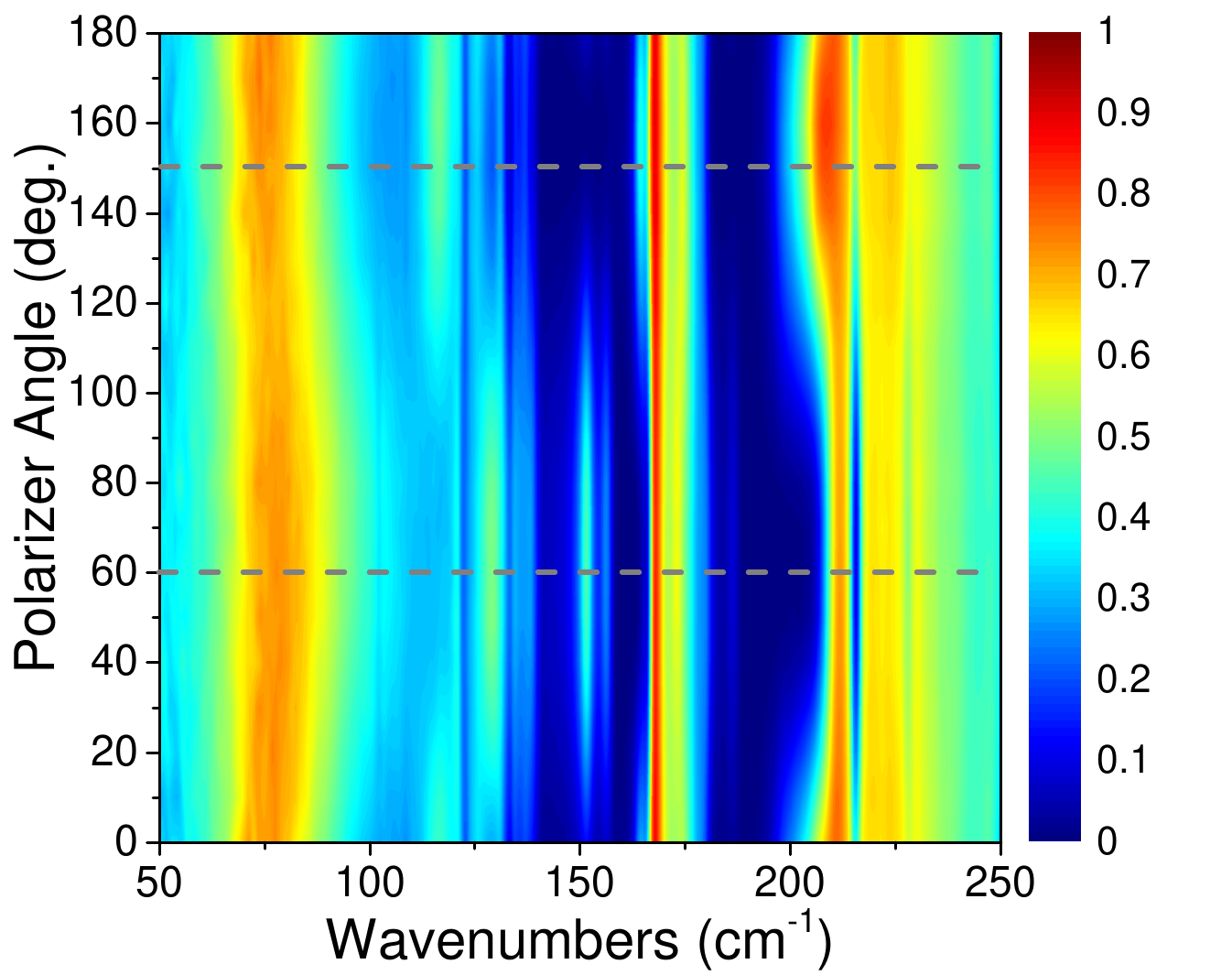}
    \caption{Colormap plot of the transmission spectra as a function of the incoming polarizer (Gold polarizer for the FIR range) angle in the range of 50 to 250\cm. The gray dashed lines represent the two principal polarizations $P$ used in this work for the temperature and magnetic field-dependent transmission measurements. Here $P=150^{\circ}$ is defined as "Pol 1" and $P=60^{\circ}$ as "Pol 2". The 122\cm excitation can be seen in this colormap as a polarization-independent absorption compared to the lattice collective modes.}
    \label{fig:poldep}
\end{figure}

As explained in the main text and in Sec.~\ref {sec:methods}, FePS$_3$ has an anisotropic response in the antiferromagnetic state. To identify the two principal optical axes for our polarization-dependent measurements, we have measured the transmission spectrum of our sample at 5~K (deep in the AFM state) and at zero field as a function of the incoming polarizer angle. Figure~\ref{fig:poldep} shows a colormap plot of the polarization-dependent transmission spectra in which two distinct spectra are identified as the principal optical axes of the FePS$_3$ single crystal sample. The same procedure was conducted in the MIR range using a KRS-5 polarizer and associated with their counterpart FIR polarized data. The entire set of transmission data from the FIR and MIR ranges was then combined into two sets termed as "Pol 1" and "Pol 2" in our work.          

\begin{figure*}[t]
    \centering
    \includegraphics[width=\linewidth]{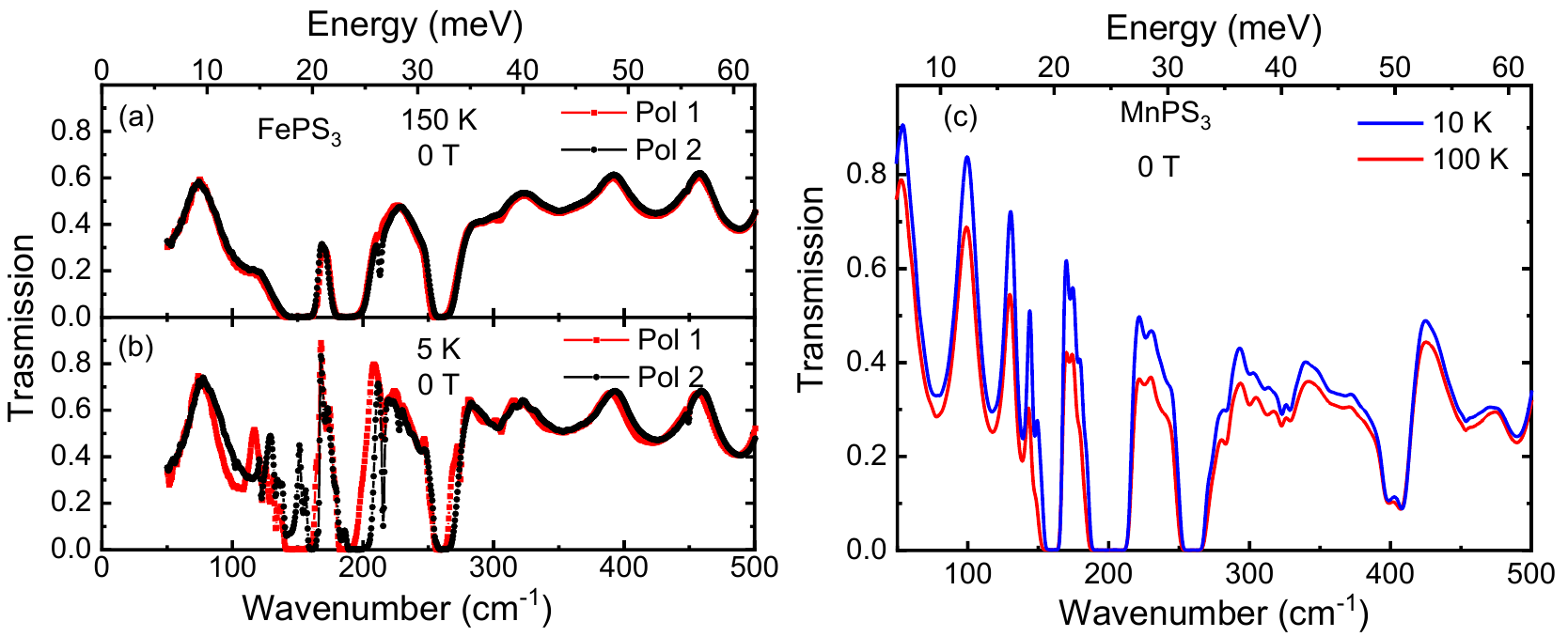}
 \caption{ Transmission spectra of FePS$_3$ ($T_N = 118$~K) for two polarizations at (a) 150~K ($T > T_N$) and (b) 5~K ($T \ll T_N$). (c) Zero-field transmission spectra at 100~K ($T > T_N$) and 10~K ($T \ll T_N$) for MnPS$_3$ ($T_N = 78$~K). No additional phonon modes emerge in the antiferromagnetic state.}
    \label{fig:Tr_comparison}
\end{figure*}

In addition to the comparison we have made in the main text, between different temperatures of the transmission spectra, we show in Figure~\ref{fig:Tr_comparison} (a)-(b) a comparison of the transmission data between the two principal polarizations at $150~K>T_N$ and at $5~K \ll T_N$, where  T$_{N}$=118~K. At 150~K, there is no significant difference between the two spectra, except for the absorption line at around 220\cm, indicating isotropic transmission behavior on the $a-b$ plane of the monoclinic structure of FePS$_3$ in the paramagnetic phase. 

\begin{figure}
    \centering
    \includegraphics[width=\linewidth]{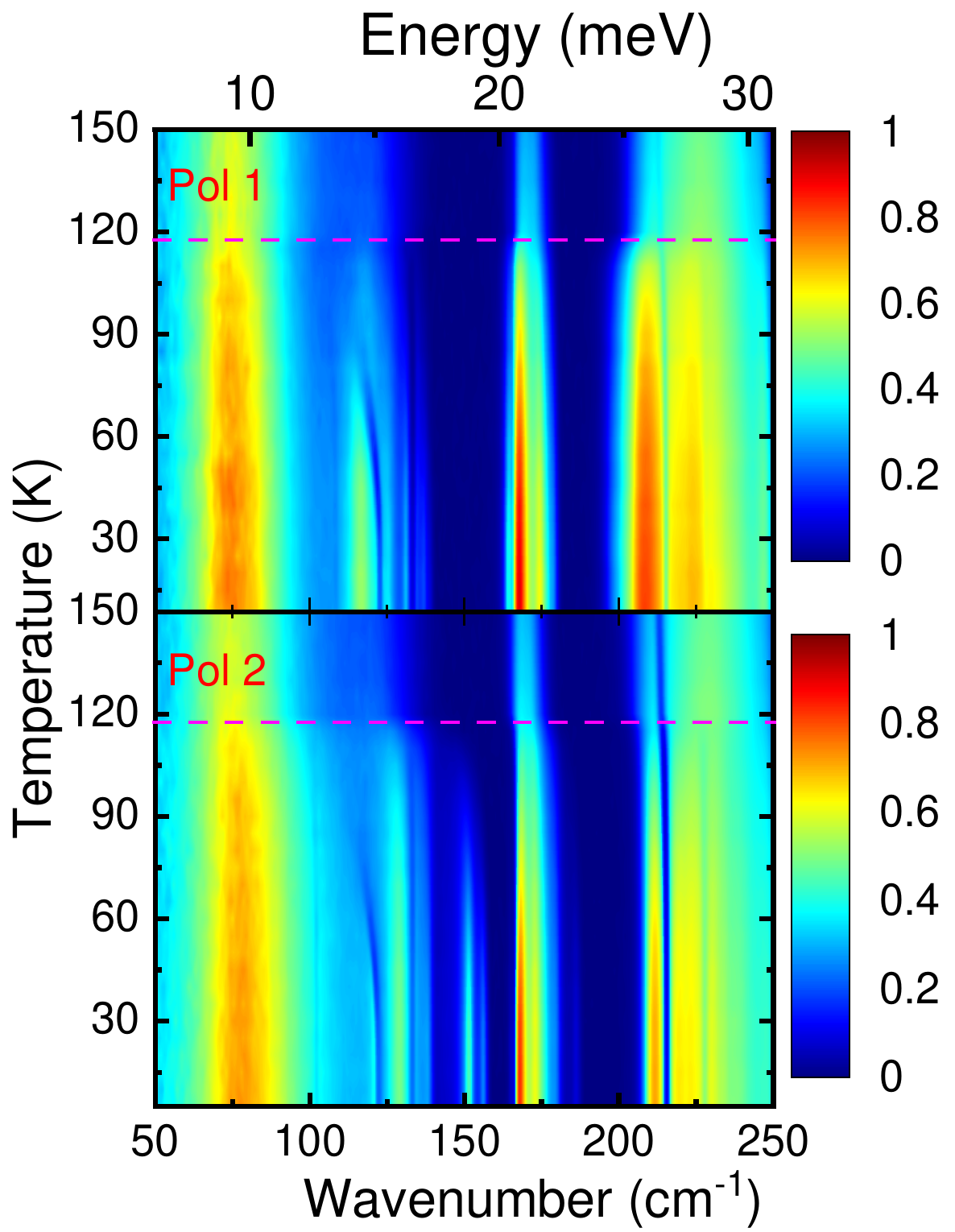}
       \caption{Polarized transmission through a 26~$\mu$m FePS$_3$ flake as a function of frequency and temperature for two principal polarizations (upper panel: Polarization 1, and lower panel: Polarization 2. See Sec .~\ref{sec:optics} for definition). A pronounced spectral evolution occurs below $T_N = 118$~K, including the emergence of new phonons and redshifting of the 122\cm (15~meV) mode.}
    \label{fig:ZF_pol}
\end{figure}

Figure~\ref{fig:ZF_pol} shows a color map of the frequency and temperature dependence of the transmission spectra in the FIR range, where deep blue colors mark areas of absorption lines due to collective excitations such as phonons. The spectra change gradually as T$_{N}$ is approached and cannot be fitted by an isotropic Drude-Lorentz model, as in the case of the high-temperature data, pointing to the onset of magnetic order and increasing spin-phonon coupling effects. 

Below $T_N$, a pronounced anisotropy develops in the spectra up to about 250~\cm, reflecting the formation of the zigzag antiferromagnetic structure in FePS$_3$. While the $C2$, mirror, and inversion symmetries of the $C2/m$ lattice are preserved, the magnetic ordering enlarges the unit cell~\cite{Bernasconi1988,Wang2016,Lee2016}, which can lead to Brillouin-zone folding and contribute to the modified phonon spectra.

To further highlight the role of the magnetic state, we compare our data with MnPS$_3$, which shares the same monoclinic crystal structure but exhibits a nearly isotropic Néel antiferromagnetic order at $T_N = 78$~K. As shown in Fig.~\ref{fig:Tr_comparison}(c), the zero-field transmission spectra of MnPS$_3$ at 100~K ($T > T_N$) and 10~K ($T \ll T_N$) reveal no additional phonon modes across the magnetic transition, consistent with previous Raman measurements~\cite{Sun2019}.

This comparison is consistent with the phonon modifications observed in FePS$_3$ not being a generic consequence of antiferromagnetic ordering in this family. Instead, they are closely linked to the anisotropic spin structure of FePS$_3$, in which ferromagnetic chains along the $a$-axis are coupled antiferromagnetically along the $b$-axis. This directional arrangement of spins leads to anisotropic exchange interactions, such that lattice vibrations modulating bonds along different crystallographic directions experience distinct effective force constants. As a result, the phonon modes undergo anisotropic renormalization through spin--phonon coupling.

Some of the weak spectral features observed in the low-energy region may originate from Brillouin-zone folding associated with the enlarged magnetic unit cell below $T_N$, as suggested in previous Raman studies~\cite{Wang2016,Lee2016,Hashemi2017,McCreary2020}. However, the pronounced polarization dependence observed in our data is consistent with anisotropic spin--phonon coupling as the dominant mechanism governing the phonon response. We note that our DFT calculations do not explicitly include magneto-elastic coupling effects, and a full treatment of the phonon spectrum in the antiferromagnetic phase remains an open theoretical challenge.

\section{Raman modes}\label{sec:appRaman}

The Raman spectroscopy measurements were performed in backscattering geometry from the $ab$ plane, without controlled alignment of the polarization axes relative to the crystallographic axes. In Table~\ref{tab:Rphononlist} we summarize the Raman modes as obtained on a FePS$_3$ single crystal at $T = 5$~K. The incoming and outgoing polarization configurations ($XX$ and $XY$) are indicated for each observed mode. We note that the assignment of irreducible representations ($A_g$, $B_g$) is not based on a strict symmetry analysis of the Raman tensor, but rather on comparison with the calculated phonon frequencies obtained from DFT.

The mode at 108\cm, which is Raman-active, exhibits a measurable Faraday rotation signal in the infrared data. The mode near 122\cm has been reported previously in both Raman and infrared measurements and is generally attributed to a magnetic excitation. In addition, several Raman-active modes show a close correspondence in frequency to weak features observed in the infrared spectra, including the modes at 161, 229, 232, and 250\cm, which are near infrared features at 160.3, 228.4, 232.2, and 251\cm, respectively. Given the experimental uncertainty in the Raman frequency calibration, on the order of $\sim$1\cm due to laser alignment, these values are consistent within experimental accuracy.

Such proximity in frequency between Raman- and infrared-active modes has been reported previously in FePS$_3$ and related compounds~\cite{Jouanne1988,Bernasconi1988}, and can arise from distinct phonon branches that lie close in energy. In this context, the observation of similar features in both spectroscopies does not imply a change in symmetry, but rather reflects the complex phonon spectrum of the material. In addition, coupling between lattice vibrations and magnetic excitations may enhance the visibility of certain modes across different experimental probes, consistent with the spin--phonon coupling discussed in the main text.

\begin{table}
\centering
\caption{Experimentally observed excitations from Raman spectroscopy of FePS$_3$ in backscattering from the $ab$ plane compared with DFT phonon calculations performed for the antiferromagnetic cell. XX and XY denote parallel and crossed polarizations of the incident and scattered light. DFT phonon symmetries and activities are taken from the calculated phonon spectrum. All values are in cm$^{-1}$.}
\begin{tabular}{|c|c|c|c|}
\hline
\textbf{Exp. freq.} & \textbf{Pol.} & \textbf{DFT freq.} & \textbf{Comment} \\
\hline

88   & XX,XY & 90.6 ($B_g$)     & Phonon on M point~\cite{Wang2016} \\
\hline

90   & --    & --               & Side phonon peak \\
\hline

95   & XX,XY & --               & Phonon on M point~\cite{Wang2016} \\
\hline

108  & XX,XY & --               & Phonon on K point~\cite{Wang2016} \\
\hline

122  & XX,XY & --               & Magnetic excitation \\
\hline

161  & XX,XY & 159.2 ($B_g$)    &  \\
\hline

191  & XY    & 188.9 ($B_g$)    &  \\
\hline

229  & XX    & 221.5 ($A_g$)    &  \\
\hline

232  & XY    & 228.5 ($B_g$)    &  \\
\hline

250  & XX,XY & 245 ($B_g$)      & \\
\hline

281  & XX    & 260.0 ($A_g$)    &  \\
\hline

283  & XY    & --               &  \\
\hline

381  & XX,XY & --               &  \\
\hline

416  & XX,XY & 415.3 ($B_g$)    &  \\
\hline

567  & XX,XY & 559.5 ($B_g$)    &  \\

\hline
\end{tabular}
\label{tab:Rphononlist}
\end{table}

\section{Electronic structure of FePS$_3$}\label{sec:CFE}
\begin{figure}[!!ht!!]
    \centering
    \includegraphics[width=\linewidth]{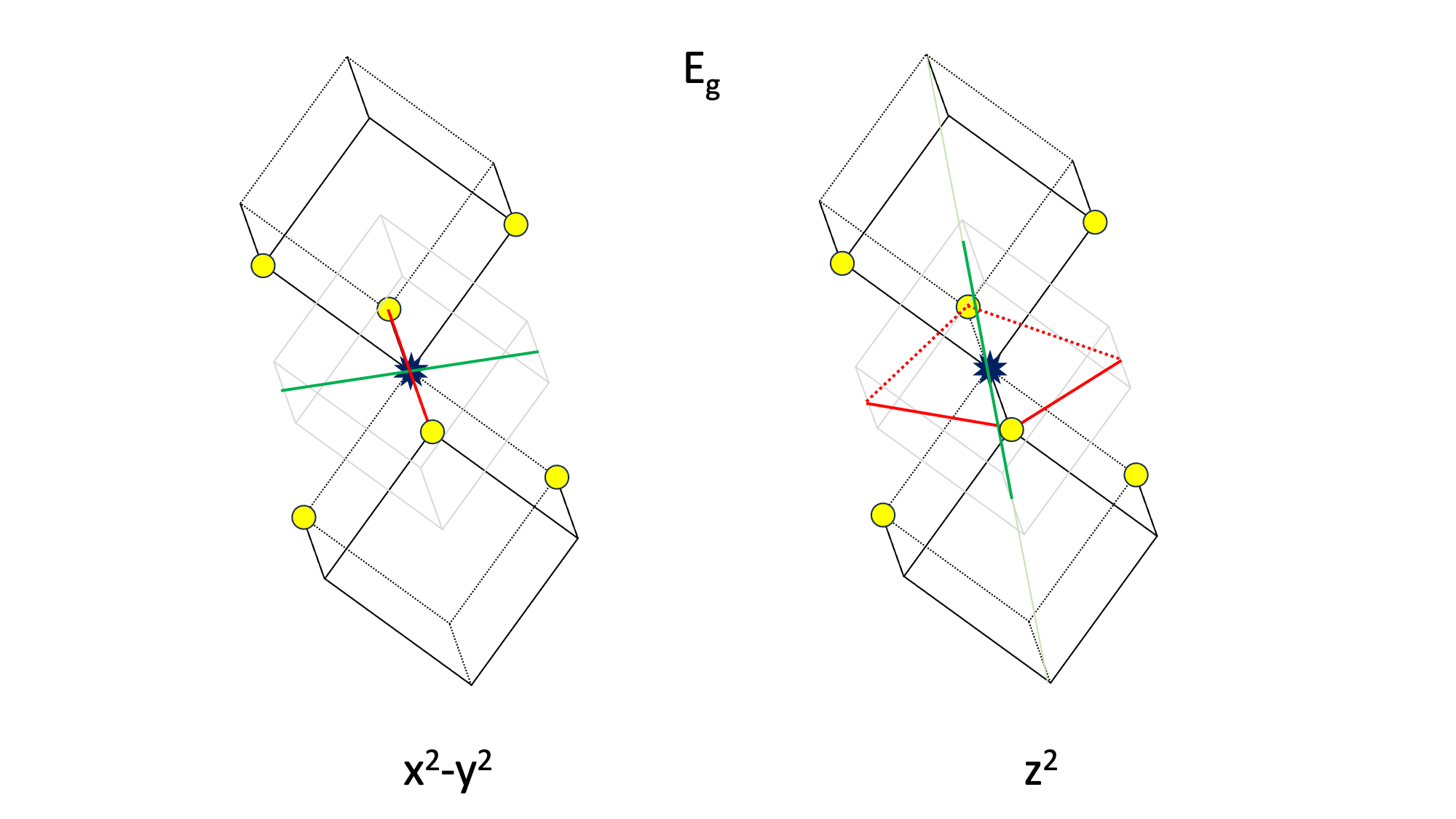}
    \includegraphics[width=\linewidth]{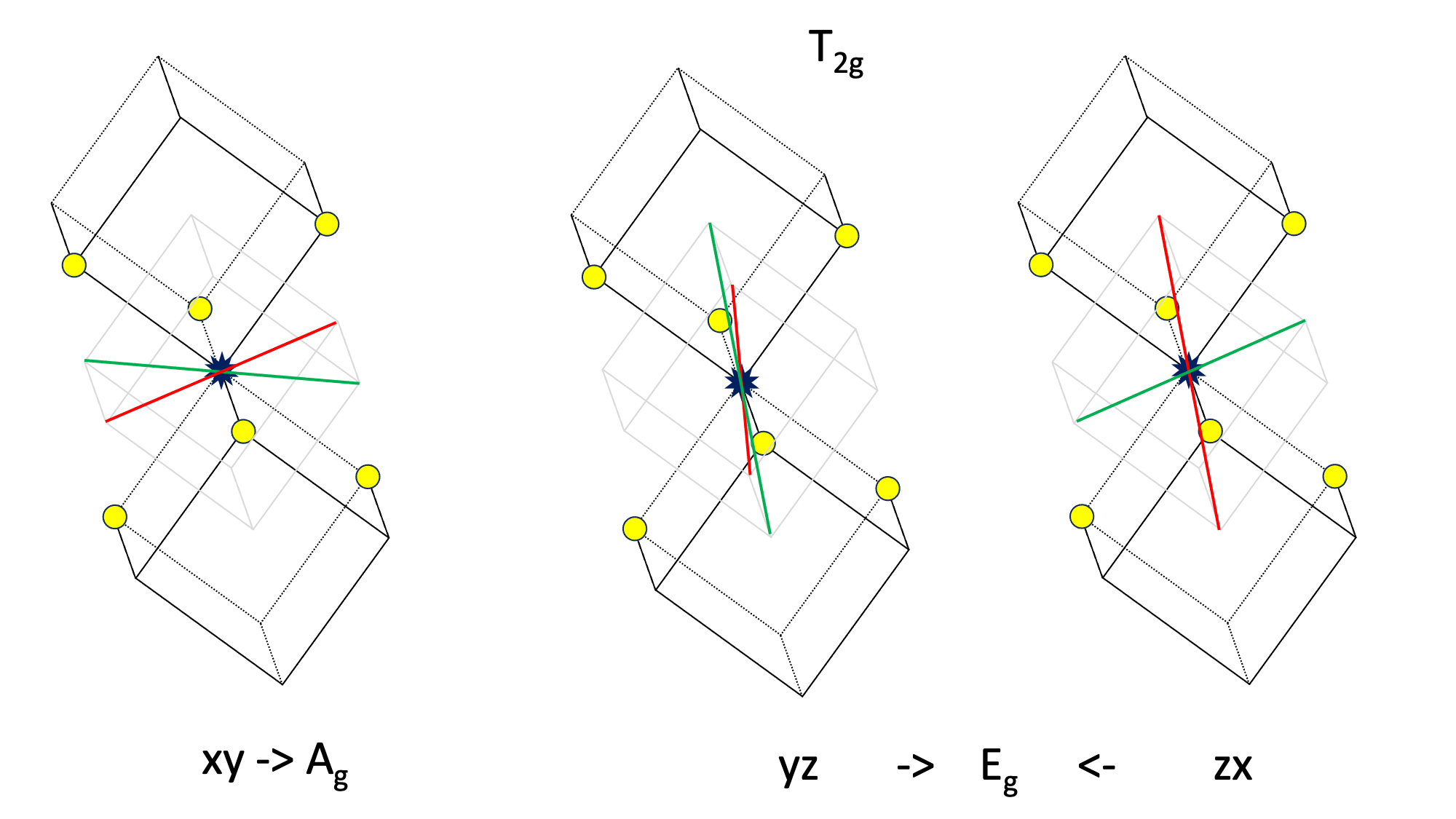}
    \caption{\label{fig:structure}
    Local crystal field environment of Fe$^{2+}$ in FePS$_3$ and orientation of the $e_g$ (top) and $t_{2g}$ (bottom) orbitals. The $e_{g}$ states have one protuberance pointing toward the negative sulfur ions and one away from it. The $t_{2g}$ states point away from the negative ions. If we call the axis connecting the farthest corners of the two cubes the $z$ axis, the $yz$ and $zx$ orbitals have one protuberance along $z$ and one protuberance in the plane. The $xy$ orbital has both protuberances pointing in the plane ({\it i.e.} away from the negative sulfur ions).}
\end{figure}
A complete understanding of the electronic structure of FePS$_3$ requires to take into account both itinerant and localized aspects. In this section we review the localized aspect starting from the assumption that the Fe atoms in FePS$_3$ 
have a $3d^6$ configuration. There exist $\binom{10}{6}=210$ different ways of filling the $5\times 2$ $3d$-states with 6 electrons. Due to the Coulomb interaction the energies of these states are spread over 10~eV and can be labeled according to spin ($S$), orbital momentum ($L$) and spin-orbital ($J$) angular momentum of the 6 electrons together. In the absence of a crystal field and spin-orbit interaction, the ground state is described by the $^5D$ multiplet (2S+1=5 and “D”=L=2) comprising $5\times 5$ degenerate eigenstates.
 
In FePS$_3$ the negatively charged sulfur ions surrounding Fe$^{2+}$ occupy roughly the 8 corners of an octahedron (see Fig.~\ref{fig:structure}). The ensuing crystal field splits the $^5D$ multiplet in $^5T_{2g}$ and $^5E_{g}$. The $^5E_{g}$ states have protuberances closer to the negative sulfur ions than the $^5T_{2g}$. Consequently, the energy of the $^5T_{2g}$ states is about 1~eV lower than that of the $^5E_{g}$ states. 

\begin{figure}[!!t!!]
    \centering
    \includegraphics[width=\linewidth]{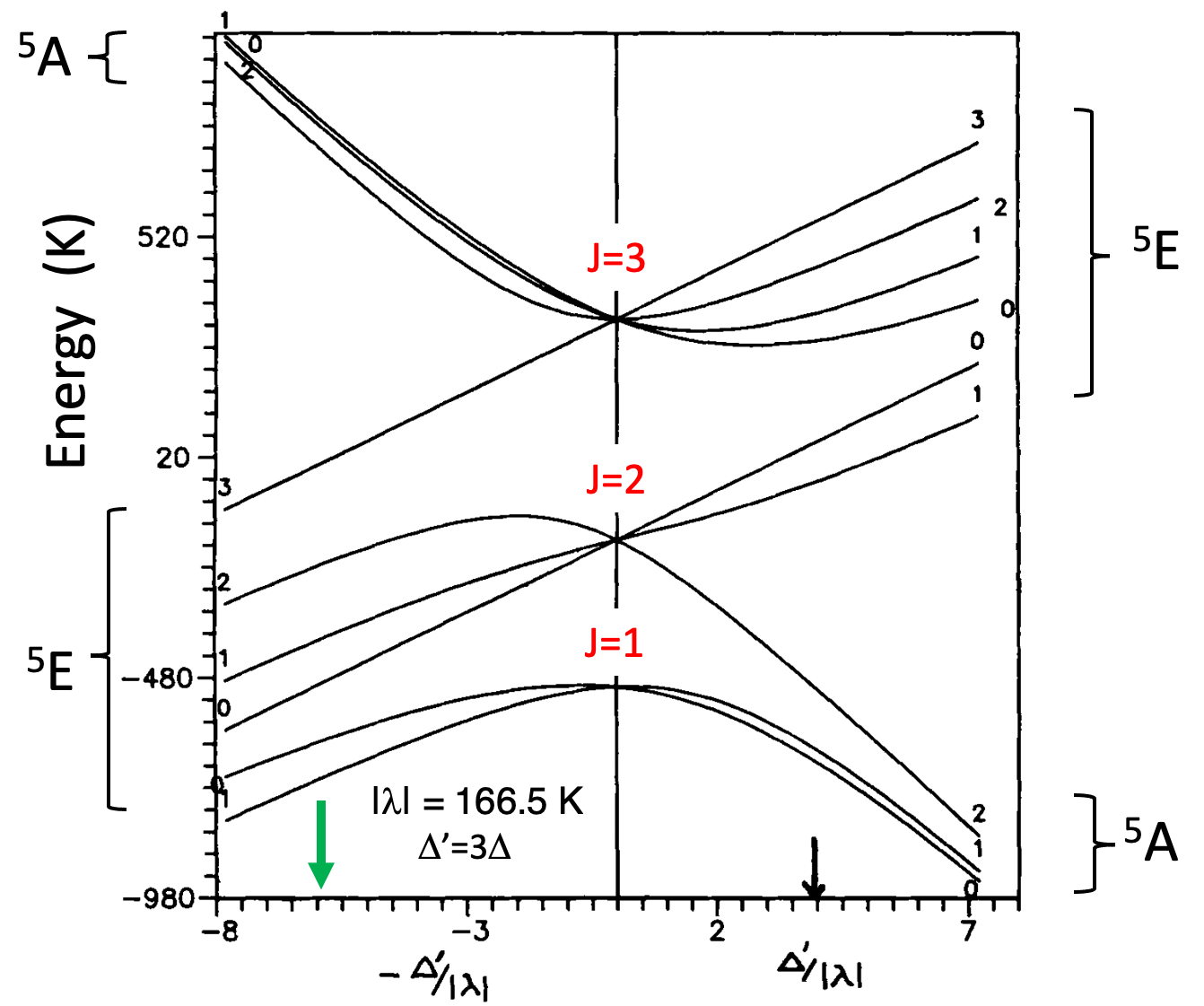}
    \caption{\label{fig:tanabe}Tanabe-Sugano diagram of the $d^6$ energy levels in a trigonal crystal field adapted from Ref.~\cite{Chandrasekharan1994}. The black arrow on the right is from the original paper. The green arrow on the left is our tentative assignment.}
    \label{fig:TanabeSugano}
\end{figure}

Spin-orbit coupling splits the $^5T_{2g}$ states in three multiplets (see Fig.~\ref{fig:tanabe}): The ground state has $J=1$, an excited $J=3$  multiplet at $\sim 30$~meV and  $J=5$ at $\sim 60$~meV.
A further refinement comes from the trigonal crystal structure, namely that two of the $^5T_{2g}$ states have one protuberance in the plane (\textit{i.e.} away from the negative sulfur ions) and one perpendicular to it. The remaining $^5T_{2g}$ state has both protuberances pointing in the plane (\textit{i.e.} away from the negative sulfur ions). The trigonal field  $\Delta$ is then negative and separates the $^5T_{2g}$ multiplet into a $^5A_g$ ground state and a $^5E_g$ excited state. Refs.~\cite{Jouanne1989,Grasso1991,Sanjun1992,Joy1992a,Joy1992b,Rule2009,Dakal2024} also obtained/assumed $\Delta<0$.  On the other hand, a positive sign was obtained in Ref.~\cite{Chandrasekharan1994} by fitting high-temperature susceptibility data, see Fig.~\ref{fig:tanabe}.
Although the orbital momentum is quenched by the crystal field, a finite orbital moment is induced by the spin-orbit coupling. This is illustrated in Fig.~\ref{fig:tanabe}, which was adapted from Ref.~\cite{Chandrasekharan1994}. The arrow on the right-hand panel is from the original paper, but in view of the expected sign of the crystal field discussed above, this seems unphysical. With this sign of crystal field the ground state of an isolated Fe$^{2+}$ ion has no magnetic moment, whereas there is a moment for $\Delta<0$. For $\Delta>0$, the possibility remains that a magnetic moment is induced by the Weiss field~\cite{Chandrasekharan1994}. 

The energy levels of our tentative assignment (green arrow in Fig.~\ref{fig:TanabeSugano}, $\Delta<0)$ are reproduced in Table~\ref{table:energies}. The highest energy states are consistent with observed absorption features in the range of 1000\cm. It is important to strike a note of caution at this point: the model used to generate this table did not take into account the orbital mixing of Fe $3d$ and the ligand states~\cite{Zaanen1985}, and this cannot be remedied by a rescaling of interaction and/or crystal field parameters because the degree of orbital mixing will in general be different from one multiplet to another according to its shape and orientation. Consequently, the energy spacings in this Table~\ref{table:energies} cannot be fully accurate. 

\begin{table}
    \begin{center}
        \begin{tabular}{c|c}
            \hline    
            Energy  (meV)   &   Degeneracy  \\
            \hline
            126.6           &   2           \\
            126.5           &   1           \\
            120.2           &   2           \\
            61.4	        &   2           \\
            41.4	        &   2           \\
            26.9            &   2           \\
            17.9            &   1           \\
            6.4             &   1           \\
            0               &   2           \\
            \hline
        \end{tabular}
        \caption{\label{table:energies} Energies of the $^5T_{2g}$-manifold split by crystal field and spin-orbit parameter, corresponding to the yellow arrow in the left panel of Fig.~\ref{fig:tanabe}}
    \end{center}
\end{table}

The ground state is a spin-orbit doublet with a finite magnetic moment. In an AF environment, the exchange field of the surrounding ions lifts the degeneracy of the ground state, making the spin point up or down. Applying a magnetic field along $z$ has the effect that for half of the spins, their energy increases, and for the other half, it decreases. If one now observes the transition from the ground state to the singly degenerate excited state at 17.9 meV, the spectrum will show a splitting of this transition, the peak moving up coming from the down-spin atoms, and the peak moving down coming from the up-spin atoms. 

Another potentially interesting feature is the transition from the ground state to the singly degenerate 6.4 meV excited state. Formally, this can be identified with the antiferromagnetic resonance~\cite {Keffer1952}. In Ref. ~\cite{Keffer1952} the phenomenon is described as the precession of the spin in an anisotropy field $H_a$ with frequency $\hbar \omega/\gamma =  H_0 \pm [H_a(2H_e+H_a)]^{1/2}$ where $H_0$ is the external field, $\gamma = ge/2mc$ and $H_e$ is the Weiss exchange field. In the local impurity case $H_a$ is replaced by the splitting of the three-fold degenerate spin-orbital state in a ground state doublet and an excited singlet state. It is important to emphasize that what we usually call the “magnon” in an optical spectrum is the lowest excitation, in table~\ref{table:energies} at 6.4 meV. As to the 122\cm absorption discussed in the present study, both the 6.4 meV or the state at 17.9 meV are possible candidates.

\section{Optical Circular Conductivity in the Birefringence Case}\label{sec:appsigma} 

We start by defining a dielectric tensor of an anisotropic sample including the off-diagonal magneto optical terms which is given by:
\begin{equation}
    \epsilon(\omega)=
    \begin{pmatrix}
        \epsilon_{xx}   &   \epsilon_{xy}   &   0               \\
        \epsilon_{yx}   &   \epsilon_{yy}   &   0               \\
        0               &   0               &   \epsilon_{zz}
    \end{pmatrix}
    \label{eq:epstensor}
\end{equation}
where $x,y,z$ are the cartesian axes. Since the light is propagating in the $z$ direction, we can refer to the in-plane dielectric tensor $e_{\perp}$, which is a $2\times2$ matrix, in the following derivation. Within this geometry, the dominant optical response is governed by the in-plane tensor components, allowing for an effective orthorhombic approximation. Therefore, the dielectric tensor in the $x-y$ plane:
\begin{equation}
    \epsilon_{\perp} = 
    \begin{pmatrix}
        \epsilon_{xx}   &   \epsilon_{xy}   \\
        -\epsilon_{xy}  &   \epsilon_{yy} 
    \end{pmatrix}
    \label{eq:epsperptensor}
\end{equation}
where the in-plane anisotropy is given by the fact that $\epsilon_{xx}\neq\epsilon_{yy}$ and the off diagonal element is $\epsilon_{yx}=-\epsilon_{xy}$.

In the same way we can define the in-plane optical conductivity tensor:
\begin{equation}
    \sigma{\perp} = 
    \begin{pmatrix}
        \sigma_{xx}   &   \sigma_{xy}   \\
        -\sigma_{xy}  &   \sigma_{yy} 
    \end{pmatrix}
    \label{eq:sigmaperptensor}
\end{equation}
since the dielectric function and optical conductivity are connected via:
\begin{equation}
    \epsilon_{\mu \nu}(\omega)=\delta_{\mu \nu}+\frac{i}{\epsilon_0\omega}\sigma_{\mu \nu}(\omega)
\end{equation}
where $\mu,\nu \in {x,y}$, $\delta_{\mu \nu}$ is the Kronecker delta, $\epsilon_0$ is the vacuum permittivity, and $\omega$ is the optical frequency.

%\subsection{Diagonalizing $\epsilon$ in the Circular Basis}

The circular polarization basis is given by the eigenvectors:
\begin{equation}
    \hat{e}_{R,L}=\frac{1}{\sqrt{2}}
    \begin{pmatrix}
        1 \\
        \mp i
    \end{pmatrix}
\end{equation}
where $L$ represents a clockwise rotation or Left-Handed (LH) circular polarization, while $R$ represents a counterclockwise rotation or Right-Handed (RH) circular polarization.

In order to convert from the $x-y$ basis to the circular basis, we need to find the matrix $U$ which will convert it as follows:
\begin{equation}
    \begin{pmatrix}
        \hat{x} \\
        \hat{y}
    \end{pmatrix}
    =
    U
    \begin{pmatrix}
        \hat{e}_R \\
        \hat{e}_L
    \end{pmatrix}
\end{equation}
which is:
\begin{equation}
    U= \frac{1}{\sqrt{2}}
    \begin{pmatrix}
        1 & 1  \\
        i & -i
    \end{pmatrix}
\end{equation}
while for the inverse case:
\begin{equation}
    \begin{pmatrix}
        \hat{e}_R \\
        \hat{e}_L
    \end{pmatrix}
    =
    U^{\dagger}
    \begin{pmatrix}
        \hat{x} \\
        \hat{y}
    \end{pmatrix}
\end{equation}
we write that:
\begin{equation}
    U^{\dagger}= \frac{1}{\sqrt{2}}
    \begin{pmatrix}
        1 & -i  \\
        1 & i
    \end{pmatrix}
\end{equation}
where $U$ is a unitary matrix and $U^\dagger U=U U^\dagger=\mathbb{1}$

We can now convert the dielectric tensor from the $xy$ basis to the circular basis as follows:
\begin{equation}
    \epsilon_{circ} = U^\dagger\epsilon_{\perp}U=
    \begin{pmatrix}
        \epsilon_{+}   &   \delta   \\
        \delta  &   \epsilon_{-} 
    \end{pmatrix}
\end{equation}
where the diagonal elements are:
\begin{equation}
    \epsilon_{\pm}=\frac{\epsilon_{xx}+\epsilon_{yy}}{2}\pm i \epsilon_{xy}
\end{equation}
and the off-diagonal elements, representing the birefringence, are:
\begin{equation}
    \delta=\frac{\epsilon_{xx}-\epsilon_{yy}}{2}
\end{equation}
Here the ($+$) sign refers to the LH circular dielectric element, $\epsilon_{+}=\epsilon_{LL}=\hat{e}^{\dagger}_L\epsilon\hat{e}_L$, the ($-$) sign refers to the RH circular dielectric element, $\epsilon_{-}=\epsilon_{RR}=\hat{e}^{\dagger}_R\epsilon\hat{e}_R$, and $\delta=\epsilon_{RL}=\epsilon_{LR}$. 

If $\delta=0$ or $\epsilon_{xx}=\epsilon_{yy}$ then the eigenmodes of the dielectric tensor are purely circular where $n_L^2=\epsilon_{+}=\epsilon_{LL}$ for LH circular polarization and $n_R^2=\epsilon_{-}=\epsilon_{RR}$ for RH circular polarization and $n=\sqrt{\epsilon}$ is the refractive index. Ref.~\citep{Levallois2015} follows this isotropic case in order to extract the Faraday (Kerr) angle rotation from the magneto-transmission (-reflection) data. However, since FePS$_3$ is anisotropic, we have to go through the entire derivation of the Faraday angle in the case of a birefringent sample.

For linear polarized light which is a superposition of LH and RH circular polarization, we can write that:
\begin{equation}
    \vec{E}_{in}=E_0\hat{x}=\frac{E_0}{\sqrt{2}}(\hat{\epsilon}_L+\hat{\epsilon}_R)
\end{equation}
The incoming field propagates through a slab of thickness $d$, where each circular component will accumulate a different phase:
\begin{equation}
    \vec{E}_{out}=\frac{E_0}{\sqrt{2}}(\hat{\epsilon}_L e^{ikn_Ld}+\hat{\epsilon}_Re^{ikn_Rd})
\end{equation}
The phase difference $\Delta\phi=\frac{\omega d}{c}(n_L-n_R)$ between the two modes will result in the Faraday rotation angle:
\begin{equation}
    \theta_F=\frac{1}{2}\Re{\left[ \frac{\omega d}{c} (n_L-n_R)\right]}
\end{equation}
and the ellipticity:
\begin{equation}
    \eta_F=\frac{1}{2}\Im{\left[ \frac{\omega d}{c} (n_L-n_R)\right]}
\end{equation}

A nonzero $\delta$ implies that the sample has a linear birefringence term and the dielectric tensor in the circular basis is not diagonal, nor $\hat{\epsilon}_R$ and $\hat{\epsilon}_L$ are the true purely circular eigenmodes. The polarization eigenstates in this case tilt away from the circular axes and mix the LH and RH components.   
    
%\subsection{Optical Circular Conductivity in the birefringent case}\label{sec:appsigma_pm}

The eigenvalues of $\epsilon_{circ}$ are therefore obtained by diagonalizing the matrix $\mathrm{det}(\epsilon_{circ}-\lambda_{\pm}\mathbb{1})=0$ but recognize that $n_{\pm}=\sqrt{\lambda_{\pm}}$, Therefore we can write the refractive indices as::
\begin{align}
    n^2_{\pm}   &= \frac{\epsilon_{+}+\epsilon_{-}}{2} \pm \sqrt{\left (\frac{\epsilon_{+}-\epsilon_{-}}{2} \right )^2+\delta^2} \\ 
                &= \frac{\epsilon_{xx}+\epsilon_{yy}}{2} \pm \sqrt{\delta^2-\epsilon_{xy}^2}
    \label{eq:npm}
\end{align}
where the first term in the right-hand side of Eq.~\ref{eq:npm} is the average of the in-plane dielectric functions:
\begin{equation}
    \bar{\epsilon} = \frac{\epsilon_{xx}+\epsilon_{yy}}{2}
    \label{eq:ebar}
\end{equation}
Below the square root are the linear birefringence term:
\begin{equation}
    \delta = \frac{\epsilon_{xx}-\epsilon_{yy}}{2}
    \label{eq:deltaeps}
\end{equation}
and the off-diagonal term which corresponds to the Magneto-Optical effect on the optical functions of the sample. Therefore we can write Eq.~\ref{eq:npm} as follows:
\begin{equation}
    \label{eq:npm2}
    n^2_{\pm} = \bar{\epsilon} \pm \sqrt{\delta^2-\epsilon_{xy}^2}
\end{equation}
We can define $\epsilon_{xx}=n_0^2 e^{+\theta}$ and $\epsilon_{yy}=n_0^2 e^{-\theta}$ with a common scale $n_0^2=\sqrt{\epsilon_{xx}\epsilon_{yy}}$ and an exponential factor $\theta=0.5 \ln(\epsilon_{xx}/\epsilon_{yy})$ in order to extract $n_{\pm}$. In fact, $n_0$ is the average refractive index or the geometric mean of $n_x=\sqrt{\epsilon_{xx}}$ and $n_y=\sqrt{\epsilon_{yy}}$, and $\theta$ represents the in-plane anisotropy, with $\theta=0$ being the isotropic case. We can then express in a different way the average of the in-plane dielectric functions as:
\begin{equation}
    \bar{\epsilon} = n_0^2\cosh{\theta}
    \label{eq:ebar_new}
\end{equation}
and the linear birefringence term as:
\begin{equation}
    \delta = n_0^2\sinh{\theta}
    \label{eq:deltaeps_new}
\end{equation}
and write Eq.~\ref{eq:npm2} as:
\begin{equation}
    \label{eq:npm3}
    n^2_{\pm} = n_0^2\left(\cosh{\theta}\pm\sqrt{\sinh^2{\theta}-\alpha^2}\right)
\end{equation}
with $\alpha=\epsilon_{xy}/n_0^2$ being a dimensionless antisymmetric parameter and much smaller than 1 for typical magneto-optical experiments. 

For the isotropic case (as in Ref.~\citep{Levallois2015}) where $\epsilon_{xx}=\epsilon_{yy}=\epsilon$ so that $\delta=0$ Eq.~\ref{eq:npm2} can be simplified to $n^2_{\pm}=\epsilon\pm\epsilon_{xy}$ and expanded to obtain:
\begin{equation}
    n_{\pm}=n_0 \pm \frac{\epsilon_{xy}}{2n_0} + \mathcal{O}(\epsilon_{xy}^2)
    \label{eq:npm_iso}
\end{equation}
with a linear in $\epsilon_{xy}$ term. The eigenmodes acquire a phase:
\begin{equation}
    \phi_{\pm}=k_{\pm}d=\frac{\omega n_{\pm}d}{c}
\end{equation}
through a slab of thickness $d$ where $c$ is the speed of light in vacuum and $k_{\pm}$ is the wavenumber for RH and LH circular polarization. Plugging in Eq.~\ref{eq:npm_iso} and $\phi_0=\omega n_0 d/c$ gives:
\begin{equation}
    \phi_{\pm}=\phi_0 \pm \frac{\omega d}{2c} \frac{\epsilon_{xy}}{n_0}
\end{equation}
The complex Faraday angle is defined as:
\begin{equation}
    \Theta = \theta_F+i\eta=\frac{1}{2}\left ( \phi_+ - \phi_- \right )
\end{equation}
and therefore allow us to write the relation between the off-diagonal terms of the dielectric tensor to the Faraday angle:
\begin{equation}
    \epsilon_{xy}(\omega)=\frac{2c}{\omega d} n_0 (\omega) \Theta(\omega)
\end{equation}
or the same for the conductivity tensor via:
\begin{equation}
    \sigma_{xy}(\omega)=-i \omega\varepsilon_0\epsilon_{xy}(\omega) 
\end{equation}
This derivation highlights the direct relation between the linear in $\epsilon_{xy}$ term to the Faraday angle rotation and the fact that the off-diagonal terms are those responsible for it.

If the crystal is slightly birefringent so that $\delta\ll\bar{\epsilon}$ the square root in Eq.~\ref{eq:npm2} can be expanded as:
\begin{equation}
    \sqrt{\delta^2-\epsilon_{xy}^2}= \delta - \frac{\epsilon_{xy}^2}{2\delta} + \mathcal{O}(\epsilon_{xy}^4)
\end{equation}
which does not have any linear in $\epsilon_{xy}$ term. The Faraday angle rotation is obtained from measurements in positive and negative magnetic fields. Therefore, the contribution from the birefringent term will be canceled out since (i) the magnetic field does not affect $\delta$ directly, (ii) $\epsilon_{xy}$ is squared and symmetric for negative and positive fields. As a result, the birefringent term will not contribute to the total Faraday angle rotation. In particular, if $\delta\ll\epsilon_{xy}$ then the rotation is still dominated by $\epsilon_{xy}$ and the circular basis is a good approximation. However, if $\delta\gg\epsilon_{xy}$ linear birefringence dominates and the Faraday effect is strongly suppressed or masked, which is not our case, as can be seen in the data.

\begin{figure}
    \centering
    \includegraphics[width=\linewidth]{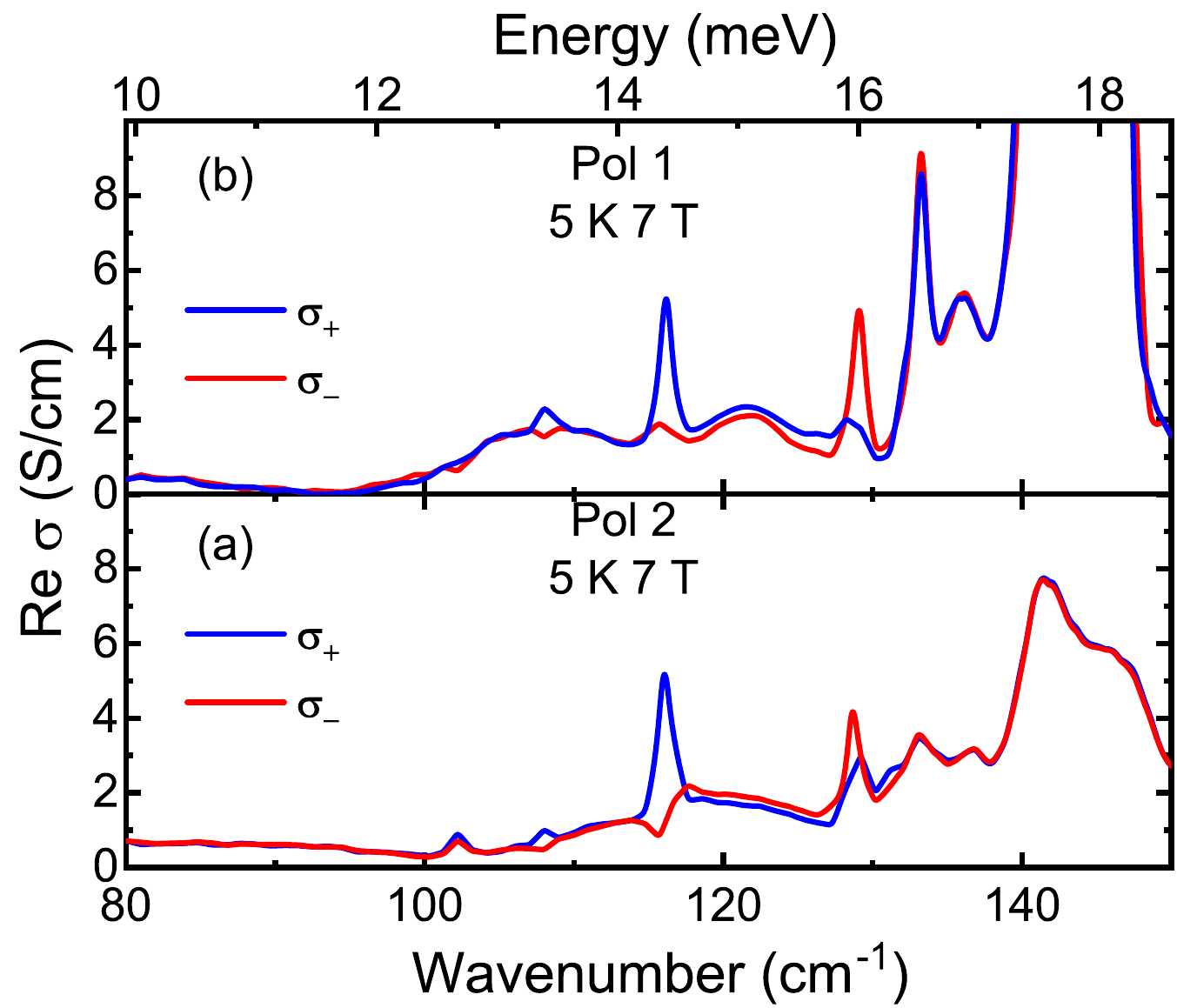}
    \includegraphics[width=\linewidth]{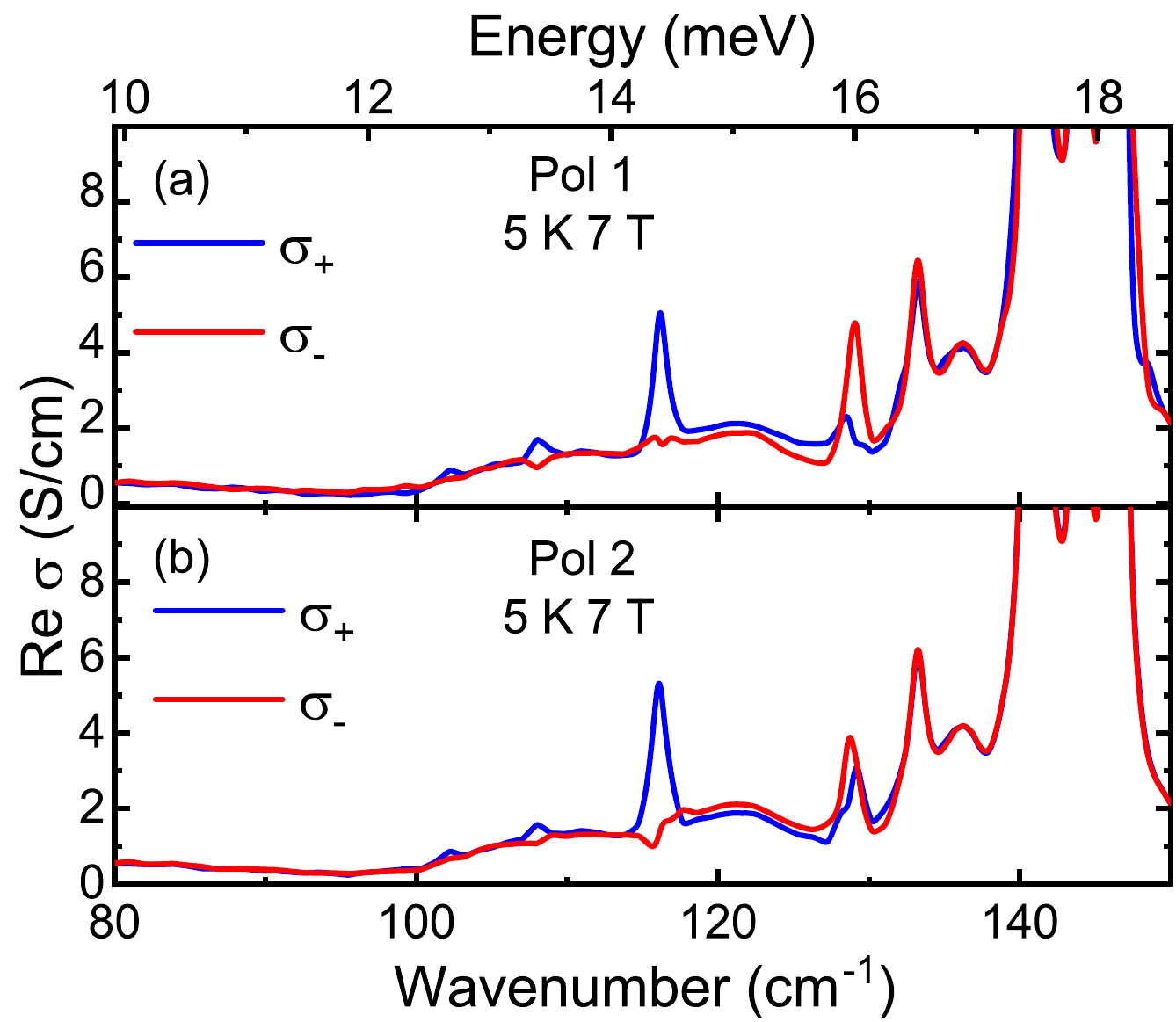}
    \caption{Top panel: Circular optical conductivities $\sigma_{+}$ and $\sigma_{-}$ reconstructed separately for each polarization (Pol~1 and Pol~2), using the combination of full-protocol Faraday rotation and absolute transmission spectra fitted using RefFit software. Bottom panel: Circular optical conductivities $\sigma_{+}$ and $\sigma_{-}$ reconstructed for each polarization (Pol~1 and Pol~2) using Eq.~\ref{Eq:sigma_pm}, where the Hall conductivity $\sigma_{xy}$ was independently derived for each polarization without averaging.}
    \label{fig:sigma_pm_aniso}
\end{figure}

To further demonstrate the negligible effect of the birefringence on the Faraday rotation, we refer to our analysis of the polarized data. The main text considers an average of the dielectric constant in order to obtain the optical circular conductivity from the Faraday angle rotation and magneto transmission measurements in the two principal axes. Figure~\ref{fig:sigma_pm_aniso} provides complementary representations of the circular optical conductivity, prior to the averaging process, which help assess the consistency of the analysis presented in the main text (Fig.~\ref{fig:sigma_pm}).

Figure~\ref{fig:sigma_pm_aniso} (top) shows $\sigma_{+}$ and $\sigma_{-}$ reconstructed separately for the two principal polarizations (noted by the two angles of the analyzer). In this case, the circular conductivities were obtained by combining the Faraday rotation measured using the full rotation protocol with the absolute transmission spectra, and fitting them directly using the MOKE + VDF procedure in RefFit software. 

Figure~\ref{fig:sigma_pm_aniso} (bottom) displays $\sigma_{+}$ and $\sigma_{-}$ for the two polarizations however now derived by using Eq.~\ref{Eq:sigma_pm} where $\sigma_{xx}$ and $\sigma_{yy}$ were averaged. Here, the off-diagonal term $\sigma_{xy}$ was determined independently for each analyzer configuration without averaging.

Fig.~\ref{fig:sigma_pm} of the main text shows the result obtained by averaging $\sigma_{xy}$ from the two polarizations, providing a single optical circular conductivity spectrum for FePS$_3$. Comparing the different plots in Fig.~\ref{fig:sigma_pm_aniso} shows that the overall important features of the circular dichroism noted in the main text, including the splitting of the 122~\cm\ mode and the spectral weight asymmetry, are reproducible across all different analysis procedures. Minor differences between the reconstructions mainly reflect systematic effects such as residual cross-polarization signals and slight misalignment of the sample.

\bibliography{refs}

\end{document}